\documentclass[12pt]{article}
\usepackage[font=small,labelfont=bf]{caption}
\usepackage{subcaption}
\usepackage{amsmath,amscd,amssymb,mathrsfs,bm}
\DeclareSymbolFontAlphabet{\mathrsfs}{rsfs}
\usepackage[mathscr]{eucal}
\usepackage{mathabx}
\usepackage{tikz}
\usepackage{tikz-cd}
\usetikzlibrary{quotes}
\usepackage{savesym}
\usepackage[nosort]{cite}
\usepackage{wasysym}
\usepackage{graphicx}
\usepackage{pifont}
\usepackage{float}
\usepackage{multicol}
\usepackage{amsfonts}
\usepackage{picture}
\usepackage{mathtools}
\savesymbol{iint}
\savesymbol{iiint}
\restoresymbol{TXF}{iint}
\restoresymbol{TXF}{iiint}
\usepackage{xcolor}
\usepackage{setspace}
\usepackage{clay}
\usepackage{slashed}
\usepackage{hyperref}
\hypersetup{colorlinks=true}
\hypersetup{linkcolor=black}
\hypersetup{citecolor=magenta}
\hypersetup{urlcolor=blue}
\usepackage{multirow}
\usepackage{verbatim}
\usepackage{accents}
\usepackage{placeins}
\usepackage{enumitem}
\usepackage[all]{xy}
\definecolor{refkey}{rgb}{0,.8,.2}\definecolor{labelkey}{rgb}{1,0,0}
\numberwithin{equation}{section}

\graphicspath{ {./Figs/} }

\newcommand{\Z}{\mathbb{Z}}
\newcommand{\R}{\mathbb{R}}

\newcommand{\C}{\mathbb{C}}

\def\Spin{\text{Spin}}

\def\Pin{\text{Pin}}

\def\tr{\text{tr}\,}

\def\dd{\text{d}}

\def\Z{\mathbb{Z}}
\def\R{\mathbb{R}}

\def\C{\mathbb{C}}

\def\sq{\text{~Sq}}

\def\mod{\text{~mod~}}

\definecolor{bittersweet}{rgb}{1.0, 0.2, 0.6}

\usepackage{array}
\newcolumntype{C}{>{$}c<{$}}

\def\RPn (#1,#2){
  \fill (#1, #2) circle (3pt);
  \fill (#1, #2+1) circle (3pt);
}

\def\sqtwoL (#1,#2,#3){
  \draw[#3] (#1,#2) .. controls (#1-1,#2+1) .. (#1,#2+2);
}

\def\sqtwoR (#1,#2,#3){
  \draw[#3] (#1,#2) .. controls (#1+1,#2+1) .. (#1,#2+2);
}

\def \sqtwoCR (#1,#2,#3){
   \draw[#3] (#1,#2) .. controls (#1+1,#2+.5) and (#1+1.5,#2+2) .. (#1+2,#2+2);
}

\def \sqtwoCL (#1,#2,#3){
   \draw[#3] (#1,#2) .. controls (#1-1,#2+.5) and (#1-1.5,#2+2)  .. (#1-2,#2+2);
}

\def \sqone (#1,#2,#3){
  \draw[#3] (#1,#2) -- (#1,#2+1);
}

\def\Aone (#1,#2){
\fill (#1, #2) circle (3pt);
\fill (#1, #2+1) circle (3pt);
\fill (#1, #2+2) circle (3pt);
\fill (#1, #2+3) circle (3pt);
\fill (#1+2, #2+3) circle (3pt);
\fill (#1+2, #2+4) circle (3pt);
\fill (#1+2, #2+5) circle (3pt);
\fill (#1+2, #2+6) circle (3pt);
\draw (#1, #2) -- (#1, #2+1);
\draw (#1, #2+2) -- (#1, #2+3);
\draw (#1+2, #2+3) -- (#1 + 2, #2+4);
\draw (#1+2, #2+5) -- (#1+2, #2+6);
\draw (#1, #2) .. controls (#1-1, #2+1) .. (#1, #2+2);
\draw (#1+2, #2+4) .. controls (#1+3, #2+5) .. (#1+2, #2+6);
\draw (#1, #2+1) .. controls (#1+1, #2+1.5) and  (#1+1.5 ,#2+3) .. (#1+2,#2+3);
\draw (#1, #2+2) .. controls (#1+1, #2+2.5) and (#1+1.5, #2+4) .. (#1+2, #2+4);
\draw (#1, #2+3) .. controls (#1+1, #2+3.5) and (#1+1.5, #2+5) .. (#1+2, #2+5);
}

\def\rectangle (#1,#2,#3){   \draw[#3] (#1-0.15,#2-0.15) rectangle (#1+0.15,#2+0.15)}

\def\Eone (#1,#2){
\fill (#1, #2) circle (3pt);
\fill (#1, #2+1) circle (3pt);
\fill (#1, #2+2) circle (3pt);
\fill (#1, #2+3) circle (3pt);
\draw (#1, #2) -- (#1, #2+1);
\draw (#1, #2+2) -- (#1, #2+3);
\draw (#1, #2) .. controls (#1-1, #2+1) .. (#1, #2+2);
\draw (#1, #2+1) .. controls (#1+1, #2+2) .. (#1, #2+3);
}

\def\joker (#1,#2){
  \foreach \y in {#2, #2+1, #2+2, #2+3, #2+4}
           {\fill (#1,\y) circle (3pt);}
           \draw (#1,#2) -- (#1, #2+1);
           \draw (#1,#2+3) -- (#1, #2+4);
           \draw (#1,#2+0) .. controls (#1-1,#2+1) .. (#1, #2+2);
           \draw (#1,#2+2) .. controls (#1-1,#2+3) .. (#1, #2+4);
           \draw (#1,#2+1) .. controls (#1+1,#2+2) .. (#1, #2+3);
}

\def\jokercolor (#1,#2, #3){
  \foreach \y in {#2, #2+1, #2+2, #2+3, #2+4}
           {\fill[#3] (#1,\y) circle (3pt);}
           \draw[#3] (#1,#2) -- (#1, #2+1);
           \draw[#3] (#1,#2+3) -- (#1, #2+4);
           \draw[#3] (#1,#2+0) .. controls (#1-1,#2+1) .. (#1, #2+2);
           \draw[#3] (#1,#2+2) .. controls (#1-1,#2+3) .. (#1, #2+4);
           \draw[#3] (#1,#2+1) .. controls (#1+1,#2+2) .. (#1, #2+3);
}

\def\msopart (#1,#2,#3){
    \fill[#3] (#1,#2) circle (3pt); 
      \fill[#3] (#1, #2+1) circle (3pt);
      \fill[#3] (#1, #2+2) circle (3pt);
      \fill[#3] (#1+2, #2+2) circle (3pt);
      \fill[#3] (#1+2, #2+3) circle (3pt);
      \fill[#3] (#1+2, #2+4) circle (3pt);
      \fill[#3] (#1+2, #2+5) circle (3pt);
    \sqtwoCR(#1,#2, #3);
    \sqtwoCR (#1, #2+1, #3);
    \sqtwoCR (#1, #2+2, #3);
    \sqone (#1+2, #2+2, #3);
    \sqone (#1+2, #2+4, #3);
    \sqtwoR(#1+2, #2+3, #3);
    \sqone (#1, #2+1, #3); }

    \def\amme (#1,#2,#3){
       \fill[#3] (#1,#2) circle (3pt) ;
   \sqtwoR(#1,#2,#3);
      \fill[#3] (#1,#2+2) circle (3pt) ;
         \sqone(#1,#2+2,#3);
      \fill[#3] (#1,#2+3) circle (3pt) ;
   \sqtwoR(#1,#2+3,#3);
   \fill[#3] (#1,#2+5) circle (3pt) ;}
   
    \def\questionupsidedon (#1,#2,#3){
       \fill[#3] (#1,#2) circle (3pt) ;
          \sqtwoR(#1,#2,#3);
                 \fill[#3] (#1,#2+2) circle (3pt) ;
             \sqone(#1,#2+2,#3);
                \fill[#3] (#1,#2+3) circle (3pt) ;}

\begin{document}

\title{Omega {\em vs.} pi, \\ and 6d anomaly cancellation}

\authors{Joe Davighi,\footnote{E-mail: {\tt jed60@cam.ac.uk}} Nakarin Lohitsiri,\footnote{E-mail: {\tt nl313@cam.ac.uk}}}
\institution{DAMTP}{\centerline{DAMTP, University of Cambridge, Wilberforce Road, Cambridge, UK}}

\abstract{
In this note we review the role of homotopy groups in determining non-perturbative (henceforth `global') gauge anomalies, in light of recent progress understanding global anomalies using bordism. We explain why non-vanishing of $\pi_d(G)$ is neither a necessary nor a sufficient condition for there being a possible global anomaly in a $d$-dimensional chiral gauge theory with gauge group $G$. To showcase the failure of sufficiency, we revisit `global anomalies' that have been previously studied in 6d gauge theories with $G=SU(2)$, $SU(3)$,  or $G_2$. Even though $\pi_6(G) \neq 0$, the bordism groups $\Omega_7^\Spin(BG)$ vanish in all three cases, implying there are no global anomalies. In the case of $G=SU(2)$ we carefully scrutinize the role of homotopy, and explain why any 7-dimensional mapping torus must be trivial from the bordism perspective. In all these 6d examples, the conditions previously thought to be necessary for global anomaly cancellation are in fact necessary conditions for the local anomalies to vanish.
}
\maketitle

\setcounter{tocdepth}{3}
\tableofcontents

\hypersetup{linkcolor=blue}

\section{Introduction}\label{sec:intro}

A common strategy for identifying global anomalies in gauge theories has been to look for non-vanishing homotopy groups. In particular, a global anomaly for gauge symmetry $G$ in $d$ dimensions has traditionally been signalled by $\pi_d(G)\neq 0$. In fact, in the usual scenario that fermions are defined using a spin structure, a global anomaly is properly characterized by the torsion part of the bordism group $\Omega^\Spin_{d+1}(BG)$, where $BG$ is the classifying space of $G$. In this note we re-examine what information is contained in $\pi_d(G)$ regarding global anomalies, and describe the ways in which $\pi_d(G)$ fails to correctly detect global anomalies. We thereby seek to reconcile previous approaches for studying global anomalies with the rigorous bordism-based criteria. 

In general there is no direct mathematical relation between $\pi_d(G)$ and  $\Omega^\Spin_{d+1}(BG)$, and so it is not surprising that there are many instances where a casual inspection of $\pi_d(G)$ might lead one to the wrong conclusion. As we discuss via several illustrative examples, the homotopy group $\pi_d(G)$ can be non-trivial while the bordism group $\Omega^{\Spin}_{d+1}(BG)$ vanishes, meaning there can be no global anomaly. This occurs, for instance, when $G=SU(2),\;SU(3)$, and $G_2$ in 6 dimensions. The reverse is also possible, in that the homotopy group can vanish despite the appropriate bordism group being non-trivial, meaning that $\pi_d(G)$ fails to detect certain global anomalies. An important example of this is given simply by any anomalous discrete gauge theory.

In order to assess the importance of $\pi_d(G)$ in this story, we must first recall why global anomalies are correctly described using bordism. Any gauge anomaly, be it local or global, always corresponds to the non-invariance of the {\em phase} of the fermionic partition function, and, fortunately, a precise formula is now known for how that phase varies under an arbitrary gauge transformation $A\to A^g$, $g(x)\in G$~\cite{Witten:2019bou}. The crucial object that appears is the $\eta$-invariant of Atiyah, Patodi, and Singer (APS)~\cite{Atiyah:1975jf,Atiyah:1976jg,Atiyah:1980jh}.\footnote{More precisely, what we call the $\eta$-invariant throughout this paper was originally introduced as the $\xi$-invariant by Atiyah, Patodi, and Singer, being the object that appears in their eponymous index theorem. \label{foot:xi}} In general, for a Euclidean spacetime manifold $M_d$,
a valid formula is always
  \begin{equation}
 Z \xrightarrow{g} Z\cdot Z_{\text{anom}}(M_d\times S^1), \qquad   Z_{\text{anom}}(M_d\times S^1) := \exp(-2\pi i \eta(M_d\times S^1)),\label{eq:anom-th}
  \end{equation}
where the $\eta$-invariant is here evaluated for an extension of the Dirac operator to a mapping torus $M_d\times S^1$, where the gauge transformation $g(x)$ is used to glue the torus along the circle direction. This formula is valid whether the gauge transformation $g(x)$ is in a trivial or non-trivial homotopy class, and so captures both local and global anomalies.
We will review this construction and this formula for the variation of the partition function in detail in \S\S \ref{sec:cobordism} and \ref{sec:trad}.

For now, it is most important to observe that the exponentiated $\eta$-invariant that appears in~\eqref{eq:anom-th} is a bordism invariant when the local anomaly polynomial vanishes. So, at least in the case when local anomalies cancel by taking the anomaly polynomial to vanish identically, there can be no global anomaly when $\Omega^\Spin_{d+1}(BG)=0$ regardless of $\pi_d(G)$. Equivalently, when $\Omega^\Spin_{d+1}(BG)$ vanishes, it is always possible to realise {\em any} mapping torus, even one whose ends are glued using a homotopically non-trivial $g(x)$, as the boundary of a bulk $(d+2)$-manifold with the gauge bundle (and spin structure) extended. Thence, the APS index theorem tells us that the anomalous phase~\eqref{eq:anom-th} is computed directly from the local anomaly polynomial, and so the most general possible anomaly must be a local one. 

Given this complete description of anomalies using bordism, it is nevertheless helpful to revisit the role of $\pi_d(G)$ in detecting global anomalies. After all, the observation that $\pi_4(SU(2)) \cong \Z/2$ played an important role in discovering the original global anomaly that afflicts $SU(2)$ gauge theory in 4d~\cite{Witten:1982fp}. In this case, if we assume that spacetime has the topology of a 4-sphere, an $SU(2)$ gauge transformation with $[g(x)]$ being non-trivial in $\pi_4(SU(2))$ can be used to construct a mapping torus $S^4 \times S^1$, as above. This mapping torus was then originally used to compute the variation of the fermionic partition function by `spectral flow' using a mod 2 version of the Atiyah--Singer index theorem~\cite{10.2307/1970757}. The mod 2 index is in fact a cobordism invariant, already hinting at the underlying importance of bordism in this calculation -- even though $\eta$-invariants did not explicitly appear in~\cite{Witten:1982fp}. But while the mapping torus is in this instance a suitable generator for the bordism group $\Omega^\Spin_{5}(BSU(2))$ (as is verified by the mod 2 index theorem), for a general theory this isn't always the case, and it is difficult to compute the bordism class of a mapping torus glued together using a homotopically non-trivial gauge transformation.

Thankfully, for the purpose of elucidating the relevance of the homotopy class of $g(x)$, there is an alternative 5-manifold we can take as a generator of $\Omega^\Spin_{5}(BSU(2))$, which is more directly linked to $\pi_4(SU(2))$. This manifold, which we will call a `mapping sphere' in the sequel, is obtained by gluing together two hemispheres using the gauge transformation $g(x)$ (see Fig.~\ref{fig:mappingsphere}), to define a 5-sphere equipped with a particular $SU(2)$ bundle. The anomaly inflow formula means that this mapping sphere is just as good a bulk 5-manifold for evaluating the global anomaly, provided that we restrict the spacetime topology to $S^4$.
Because of the natural isomorphism $\pi_4(SU(2)) \cong \pi_5(BSU(2))$, the fact that $[g(x)] \neq 0 \in \pi_4(SU(2))$ means that the bundle on the mapping sphere cannot be extended to a bulk 6-ball whose boundary is the $S^5$. And in fact, in this instance, the bundle on the mapping sphere cannot be extended to {\em any} bulk 6-manifold whatsoever, and so the APS index theorem cannot be used to compute the anomaly using the local anomaly polynomial - it is a genuine global anomaly.
Bordism invariance means, of course, that we get the same result for the variation of the partition function whether we use the mapping sphere or the mapping torus -- with the crucial difference that the mapping torus works for any spacetime $M_d$, while using the mapping sphere requires fixing $M_d=S^d$.

This example is highly instructive. In the general case, 
the non-vanishing of $\pi_d(G)$ implies, at least when $G$ is connected,\footnote{This caveat already reveals the failure of $\pi_d(G)$ to detect an important class of global anomalies, namely those occurring when $G$ is a discrete group. In this case, one cannot even write a gauge transformation in the form $g:S^d\to G$. Rather, a gauge transformation for discrete $G$ should be specified as a change of local trivialisation of the corresponding principal $G$-bundle over spacetime.  Homotopy groups thus do not probe anomalies associated with such a gauge transformation (which are necessarily global anomalies because there are no infinitesimal discrete gauge transformations).} 
and only when spacetime has the topology of $S^d$, that the variation of the partition function is computed by evaluating $\eta$ on a mapping sphere $S^{d+1}$ whose $G$-bundle cannot be extended to a bulk $(d+2)$-ball, where the $S^{d+1}$ is obtained by gluing together two hemispheres using the homotopically non-trivial gauge transformation. But $\pi_d(G)$ tells us nothing about the existence of other (non-contractible) bulk manifolds bounded by $S^{d+1}$, and the existence of such a manifold would mean there is no global anomaly.  The more difficult question of whether {\em any} such bulk exists can only be answered using bordism theory. This is a precise way of seeing why $\pi_d(G) \neq 0$ is not sufficient for a $d$-dimensional $G$-gauge theory to have a global anomaly, even when spacetime is taken to be a sphere. Nonetheless, we see from the mapping sphere construction that it {\em is} true that $\pi_d(G) \neq 0$ is {\em necessary} for this theory to have a global anomaly when restricted to $M_d = S^d$, and when $G$ is connected, because if $\pi_d(G)=0$ then the mapping sphere can always be filled in with a $(d+2)$-ball.

We back up these general arguments by considering some well-known examples in 6 dimensions, where the conflict between $\Omega^\Spin_{d+1}$ and $\pi_d$ can be seen especially clearly. In particular, we carefully analyze a 6d gauge theory with gauge group $SU(2)$.  In this case, a global anomaly is an `obstruction' to extending both the $SU(2)$ gauge configuration and the spin structure from any suitable mapping torus (in a general sense) that implements the supposedly globally anomalous gauge transformation in $\pi_6(SU(2))\cong \Z/12$, to {\em any} suitable spin 8-manifold that it bounds. We emphasize that $\pi_6(SU(2)) \neq 0$ only obstructs a particular extension of the gauge field, namely that from a 7-dimensional  mapping sphere, as described above, 
into an 8-ball, and takes no account of the spin structure. (One gets away with ignoring the spin structure in this special case, because the unique spin structure on $S^d$ always extends when $d>1$). More properly, one should conjointly consider the homology groups of $BSU(2)$ and the spin bordism groups of a point if one searches for a generic extension with arbitrary topology, and for an original spacetime also of arbitrary topology. The correct way to combine these pieces of information is via the Atiyah--Hirzebruch spectral sequence, which computes the spin bordism groups of $BG$. In this instance, computing that $\Omega^\Spin_{7}(BSU(2))=0$ is a shortcut to showing that {\em any} generalized mapping torus must be null-bordant and so there can be no global anomaly.

One might wonder what happens when the anomaly polynomial does not vanish identically, but rather the various perturbative anomalies are cancelled by the Green--Schwarz mechanism, as occurs in 6d $\mathcal{N}=(1,0)$ supergravity theories. It turns out that very little of our preceding discussion changes. It was recently established~\cite{Monnier:2018nfs} that a `non-perturbative' generalization of the Green--Schwarz term, in the form of a \emph{shifted Wu--Chern–Simons theory}~\cite{Monnier:2016jlo}, always differs from the anomaly theory \eqref{eq:anom-th} by a bordism invariant.  If we denote the Green--Schwarz anomaly theory that cancels the local anomalies by $Z_{\text{GS}}$, then the new partition function $\tilde{Z}$, modified by the appropriate Green--Schwarz term, transforms as $\tilde{Z} \rightarrow \tilde{Z} \cdot Z_{\text{glob}}$ under any (possibly homotopically non-trivial) gauge transformation $g(x)$, where $Z_{\text{glob}} = Z_{\text{anom}}\cdot Z_{\text{GS}}$ and is a bordism invariant. In particular, if $\Omega^\Spin_7(BG) = 0$, then the Green--Schwarz-shifted partition function $\tilde{Z}$ remains invariant under \emph{any} gauge transformation, regardless of $\pi_6(G)$.

Monnier and Moore~\cite{Monnier:2018nfs} went on to establish that there can be no global anomaly in 6d for gauge groups $G=SU(n), Sp(n), U(n), E_8$ or any arbitrary products between them, by explicitly showing that $\Omega^\Spin_7(BG)$ vanishes. We add to their analysis by computing explicitly that $\Omega^\Spin_7(BG_2) = 0$ for the exceptional Lie group $G_2$, a group for which a global anomaly was previously reported in the literature (see {\em e.g.}~\cite{Kiritsis:1986mf,Bershadsky:1997sb,Dobrescu:2001ae,Suzuki:2005vu}). In summary, there can be no 6d global anomalies for any of the groups $SU(2),\; SU(3)$, or $G_2$, despite their non-trivial 6\textsuperscript{th} homotopy groups.

The rest of the paper is organised as follows. In \S\ref{sec:cobordism} we recall how global anomalies are properly characterized by bordism invariants. While this material is not new, our purpose here is to offer a pedagogical explanation of the bordism classification of global anomalies, intended to be accessible for model builders (working in various dimensions) concerned with anomaly cancellation. In \S\ref{sec:trad}, we examine the traditional global anomalies in light of this bordism perspective, contrasting the use of mapping tori and mapping spheres. In \S\ref{sec:6dexamples} we focus on examples from 6d gauge theories. We examine the homotopy-inspired mapping torus with gauge group $SU(2)$ in detail, and discuss the role of the Green--Schwarz mechanism following Monnier and Moore. A new bordism calculation of $\Omega^\Spin_7(BG_2)$ using the Adams spectral sequence is included in Appendix~\ref{app:G2bord}.

\paragraph{Note added:} while we were completing this paper we received a draft of Ref.\cite{Lee:2020} from Y. Lee and Y. Tachikawa, which investigates many of the questions explored here from a similar perspective.

\section{Anomalies, inflow, and locality}\label{sec:cobordism}

In this Section, we review why anomalies that arise from integrating out chiral fermions are correctly detected by bordism groups. This relies on the paradigm of anomaly inflow.

We consider a generic theory of chiral fermions defined on a closed Euclidean $d$-manifold $M_d$ with symmetry $G$, a compact Lie group. In particular, consider a set of Weyl fermions, denoted collectively by $\psi$, in some representation ${\bf R}$ of $G$, and which are defined (for simplicity) using a spin structure $\sigma$. To find out whether such a theory is anomalous, one couples the fermions $\psi$ to a background gauge field $A$ for $G$ via a self-adjoint Dirac operator $i\slashed{D} = \sum_{\mu=1}^d \gamma^\mu D_\mu$. The choice of principal $G$-bundle over $M_d$, on which $A$ is a choice of connection, corresponds to a map $f:M_d \to BG$. The result of integrating over the chiral fermions $\psi$ is called the fermionic partition function, henceforth denoted by $Z[A,M_d]$, where here `$A$' denotes all relevant background fields. This is formally equal to the product of eigenvalues of $i\slashed{D}^+$, the chiral counterpart of $i\slashed{D}$, in other words 
$$Z[A,M_d] = \mathrm{Det}(i\slashed{D}^+).$$
The eigenvalues of $i\slashed{D}$ on closed manifolds always come in pairs of positive and negative signs, only one of which should be chosen to contribute as an eigenvalue of $i\slashed{D}^+$.  This is the source of a sign ambiguity in defining $\mathrm{Det}(i\slashed{D}^+)$. All anomalies can be traced back to this sign ambiguity (though anomalies in general are not just a mod 2 effect due to regularisation of the determinant, $i\slashed{D}$ being an operator on an infinite-dimensional vector space).\footnote{Note that when there is no charge conjugation symmetry in the theory, one should generalize the Dirac operator to an antisymmetric operator and work with its Pfaffian instead. We will not need to discuss such intricacies further in this paper.} 

The $G$-symmetry is non-anomalous if the determinant $Z[A,M_d]$ is a genuine $\C$-valued function on the space $\mathscr{X}$ of background data, in this case on the space of connections $A$ modulo gauge transformations. This is of course not always the case. Generally, $Z[A,M_d]$ is only guaranteed to be a section of a complex line bundle over $\mathscr{X}$. If there exists a gauge transformation $A \to A^g$ which acts non-trivially on that section, then the theory is anomalous.
In fact, the modulus of the fermionic partition function is necessarily anomaly-free, and at most the gauge transformation $A \to A^g$ can result in  $Z[A,M_d]$ changing by a phase.

Thus, to understand when there is an anomaly, one needs to understand the phase of the fermionic partition function, and whether it is well-defined with respect to gauge transformations. This phase can be understood using anomaly inflow. At this point, it is convenient to divide our attention into two kinds of anomaly. A {\em local anomaly} in $G$ occurs when $Z[A+\delta_\lambda A] \neq Z[A]$ for an infinitesimal gauge transformation $\delta_\lambda A = \dd \lambda + i[\lambda,A]$, even after adjusting all possible counterterms. Such a local anomaly can be computed by one-loop diagrams in perturbation theory, generalizing the pioneering calculations of Adler, Bell, and Jackiw (ABJ)~\cite{PhysRev.177.2426,Bell:1969ts}. On the other hand, a {\em global} anomaly is one that cannot be seen in perturbation theory~\cite{Witten:1982fp}. A global anomaly will always be associated to gauge transformations that cannot be obtained from successive infinitesimal gauge transformations, and so cancellation of local anomalies does not rule out the possibility of a non-perturbative global anomaly. 

\subsection*{Anomaly inflow: perturbative version}

Precisely, the local anomaly can be written in terms of the variation of the phase of the fermionic partition function, {\em viz.}
\begin{equation}
  \label{eq:localanom}
  Z[A+\delta_\lambda A,M_d] = Z[A,M_d] \exp\left(-2\pi i \int_{M_d} I_{d}(A,\lambda) \right),
\end{equation}
for a certain $d$-form $I_{d}(A,\lambda)$. For an example that is most familiar to high energy physicists, take $d=4$ and $G=U(1)$ such that $\delta_\lambda A = \dd \lambda$, and let $\psi$ denote a single Weyl fermion of unit charge. Then the calculations of ABJ (see also~\cite{PhysRev.182.1517,PhysRevLett.42.1195}) tell us that $I_4(A,\lambda) = \frac{\lambda}{8\pi^2} F \wedge F$, where $F=dA$ is the field strength. 

The idea of anomaly inflow is that this variation in the phase of the fermionic partition function can be precisely reproduced by the variation of a classical Chern--Simons action on a 5-dimensional bulk $X$ whose boundary is $M_4$~\cite{Jackiw:1983nv,Zumino:1983ew,Stora:1983ct,Callan:1984sa}, and to which we have extended the connection $A$, the spin structure $\sigma$, and thence the fermion field.\footnote{Such extensions will not always exist. This is the case even when there is no gauge group, since the K3 surface generates $\Omega^\Spin_4(\text{pt})\cong \Z$ meaning that its spin structure cannot be extended to a bulk 5-manifold. The possibility for such `non-nullbordant' spacetimes does not in fact give rise to any further anomalies, but does give rise to a real ambiguity in defining the partition function. This is reflected in a choice of gravitational theta-angle~\cite{Freed:2004yc}.} Specifically, the partition function for the classical action is a phase $\exp(-2\pi iS_{\text{CS}}) = \exp\left(-2\pi i\int_X I_5(A)\right)$, where $I_5(A)= \frac{1}{8\pi^2} A \wedge F \wedge F$ is the Chern--Simon 5-form. This Chern--Simons form is not gauge-invariant on a 5-manifold with boundary but rather shifts precisely by the phase in \eqref{eq:localanom}. In fact, the original anomalous 4d partition function can be written as $Z[A,M_4]=|Z[A,M_4]|\exp\left(-2\pi i\int_X I_5(A)\right)$, where recall the modulus $|Z[A,M_4]|$ is anomaly free. The version of this story in two dimensions lower is especially familiar in condensed matter physics, because it describes the integer quantum Hall effect.

For a general local anomaly, the $d$-form $I_{d}(A,\lambda)$ is related to the anomaly polynomial $\Phi_{d+2}(A)$ by the descent equations~\cite{AlvarezGaume:1983ig,AlvarezGaume:1984dr,Bardeen:1984pm}
\begin{equation}
  \label{eq:anomdesc}
  \Phi_{d+2}(A) = \dd I_{d+1}(A), \quad I_{d+1}(A,\lambda) = \delta_\lambda I_{d+1}(A), \quad I_{d+1}(A,\lambda) = \dd I_{d}(A,\lambda).
\end{equation}
Here $\Phi_{d+2}(A)$ is itself a closed, gauge-invariant $(d+2)$-form that will play a central role in what follows, given by
\begin{equation}
  \label{eq:anom-poly}
  \Phi_{d+2}(A;{\bf R}) = \hat{A}\,\tr_{{\bf R}} \exp\left(\frac{F}{2\pi}\right)\bigg\rvert_{d+2},
\end{equation}
where $\hat{A}$ is the Dirac genus of the tangent bundle and $F$ denotes the curvature of the connection $A$. The bar and subscript `$d+2$' indicates that one should take only the $(d+2)$-form terms on the right-hand side. Provided that the gauge field $A$ and spin structure $\sigma$ can be extended from $M_d$ to a bounding $(d+1)$-dimensional bulk manifold $X$, the phase of the (possibly anomalous) fermionic partition function is given precisely by anomaly inflow, {\em viz.}
\begin{equation} \label{eq:inflow_pert}
Z[A,M_d]=|Z[A,M_d]|\exp\left(-2\pi i\int_X I_{d+1}(A)\right).
\end{equation}
The local anomaly cancels if and only if $\Phi_{d+2}(A) = 0$.

\subsection*{Anomaly inflow: non-perturbative version}

The anomaly inflow paradigm can also be used to describe non-perturbative global anomalies. In the rest of this Section (and the next), we aim to give a pedagogical review of chiral fermion anomalies from the anomaly inflow perspective, which is crucial to understanding why global anomalies are correctly detected by bordism groups -- and thus why the non-vanishing (or not) of homotopy groups at best contains only partial information about the global anomaly.

The starting point is a generalization of the formula~\eqref{eq:inflow_pert} for the phase of the fermionic partition function. By integrating out massive fermions in the bulk manifold $X$ (where $\partial X = M_d$), 
the following formula for the fermionic partition function can be obtained~\cite{Witten:2019bou},
\begin{equation} \label{eq:inflow_Npert}
Z[A,M_d]=|\mathrm{Det}(i\slashed{D}_{M_d}^+)|\exp\left(-2\pi i\eta(X) \right),
\end{equation}
where here $\eta(X)$ denotes the Atiyah--Patodi--Singer (APS) $\eta$-invariant~\cite{Atiyah:1975jf,Atiyah:1976jg,Atiyah:1980jh} associated to an extension of the Dirac operator from $M_d$ to the bulk manifold $X$ using APS boundary conditions. The $\eta$-invariant is a regularised sum of the signs of the eigenvalues $\lambda_k$ of the Dirac operator on $X$. One possible regularization is\footnote{We assume the convention that any zero modes of $i\slashed{D}_X$ are counted with a positive sign, {\em viz.} $\text{sign}(0)=+1$. As mentioned in footnote \ref{foot:xi}, the resulting `$\eta$-invariant' defined in (\ref{eq:eta}) in fact corresponds to the $\xi$-invariant in the original notation of Atiyah, Patodi, and Singer~\cite{Atiyah:1975jf,Atiyah:1976jg,Atiyah:1980jh}.
}
\begin{equation} \label{eq:eta}
\eta(X) =
\text{lim}_{\epsilon\to 0^+} \sum_k e^{ -\epsilon |\lambda_k|}\text{sign}(\lambda_k)/2.
\end{equation}
The formula~\eqref{eq:inflow_Npert} contains full information about the local and global anomalies that can afflict the original $d$-dimensional theory on $M_d$, as we will soon review. The role of the $\eta$-invariant in capturing global anomalies was first appreciated in~\cite{Witten:1985xe} (see also~\cite{Witten:2015aba}).

The anomaly inflow formulae (\ref{eq:inflow_pert}, \ref{eq:inflow_Npert}) will no doubt strike the unfamiliar reader as problematic descriptions of a theory on $M_d$, because there is seemingly a dependence on extending the theory to an unphysical bulk manifold $X$. This involves many choices, such as how to extend the gauge field $A$ and the spin structure $\sigma$ to $X$. Any dependence of $Z[A,M_d]$ on these choices will signify a sickness of the $d$-dimensional theory, which amounts to an anomaly. From this point of view, the anomaly can be understood as a failure of locality, manifest in a dependence of supposedly physical quantities on the unphysical bulk $X$. 

\begin{figure}[h]
\centering
\begin{subfigure}[b]{0.85\textwidth}
  \centering
   \includegraphics[width=0.7\textwidth]{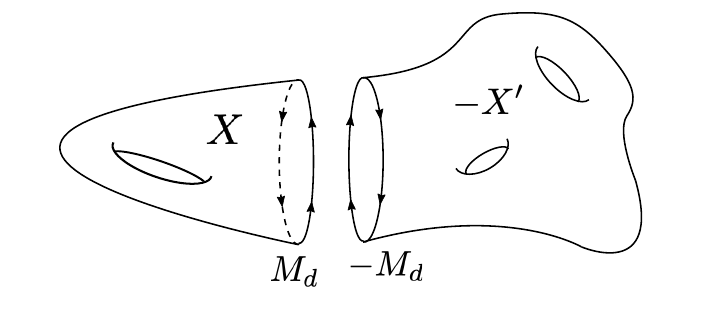}
 \end{subfigure}

\begin{subfigure}[b]{0.85\textwidth}
  \centering
   \includegraphics[width=0.7\textwidth]{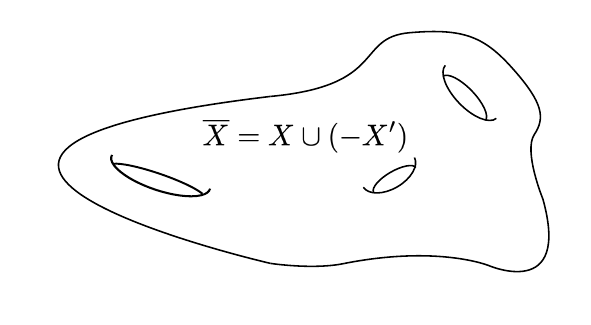}
\end{subfigure}
\caption{Gluing of two manifolds $X$ and $X^\prime$ with a shared boundary component $M_d$, under which the exponentiated $\eta$-invariant factorizes due to a theorem of Dai and Freed~\cite{Dai:1994kq}.} \label{fig:glue}
\end{figure}

We are now in a position to ask when~\eqref{eq:inflow_Npert}, which is a fully non-perturbative formula for the fermionic partition function, defines a theory consistent with locality. Suppose we find an alternative extension of the theory to a different bulk manifold $X^\prime$, equipped with different choices of extension for the gauge field and spin structure. Locality requires that the partition function is independent of these choices, thus
\begin{equation} \label{eq:loc_begin}
\exp\left(-2\pi i\eta(X) \right) = \exp\left(-2\pi i\eta({X^\prime}) \right) .
\end{equation}
This equation can be usefully rearranged thanks to a ``gluing formula'' for the exponentiated $\eta$-invariant~\cite{Dai:1994kq}, as follows. Since $X$ and $X^\prime$ share the same boundary, $\partial X = \partial X^\prime = M_d$, one can glue them together along this shared boundary if we first `flip' the orientation of, say, $X^\prime$. Denoting this orientation-reversed copy of $X^\prime$ by $-X^\prime$, the result of the gluing is a closed $(d+1)$-dimensional manifold $\overline{X} := X \cup (-X^\prime)$, 
as illustrated in Fig.~\ref{fig:glue}. 

The gluing formula is that~\cite{Dai:1994kq}
\begin{equation} \label{eq:glue}
\exp\left(-2\pi i\eta(X) \right) \exp\left(-2\pi i\eta(-X^\prime) \right) = \exp\left(-2\pi i\eta(\overline{X}) \right).
\end{equation}
Thus, using also $\exp\left(-2\pi i\eta(-X^\prime) \right) = \exp\left(+2\pi i\eta(X^\prime) \right)$, the condition~\eqref{eq:loc_begin} for locality becomes simply
\begin{equation} \label{eq:locality}
\exp\left(-2\pi i\eta(\overline{X}) \right) = 1 .
\end{equation}
Indeed, to be fully consistent with locality, the triviality of the exponentiated $\eta$-invariant~\eqref{eq:locality} should hold not just for the particular manifold $\overline{X}$, but for {\em all} closed $(d+1)$-manifolds equipped with the necessary structures. This condition will imply cancellation of all known local and global anomalies. 

To see the power of the locality condition~\eqref{eq:locality}, it is helpful to first recall the APS index theorem~\cite{Atiyah:1975jf,Atiyah:1976jg,Atiyah:1980jh}. For a self-adjoint Dirac operator $\mathcal{D}_Y:=i\slashed{D}_Y$ on a manifold $Y$ in dimension $d+2$, whose boundary is a closed $(d+1)$-manifold $\partial Y = \overline{X}$ and where $i\slashed{D}_Y$ is defined with APS boundary conditions, the index theorem is that
\begin{equation} \label{eq:APS}
\mathrm{Ind}\left(\mathcal{D}_Y\right) = \int_Y \Phi_{d+2} - \eta(\overline{X}),
\end{equation}
where as usual $\mathrm{Ind}\left( \mathcal{D}_Y \right)=\mathrm{Dim~}\mathrm{ker~}\mathcal{D}^+_Y - \mathrm{Dim~}\mathrm{ker~}\mathcal{D}^-_Y$, and where $\Phi_{d+2}$ is the same anomaly polynomial defined in~\eqref{eq:anom-poly}. Thus we see that the $\eta$-invariant provides a boundary correction to the original Atiyah--Singer index formula.\footnote{Note that, due to this boundary correction, the APS formula implies that the index can be non-zero in even or odd dimensions when $Y$ has a boundary (whereas when $Y$ has no boundary, the Atiyah--Singer formula implies the index vanishes whenever $d$ is odd).}

Let us discuss some consequences of this index theorem. Firstly, in perturbation theory we can assume that the closed $(d+1)$-manifold $\overline{X}$ that appears in our locality condition~\eqref{eq:locality} is always the boundary of a manifold $Y$ in one dimension higher (see {\em e.g.}~\cite{Witten:2019bou}), with all the requisite structures suitably extended, such that we can directly apply the APS index theorem to evaluate the exponentiated $\eta$-invariant. Because the index is an integer, this yields
\begin{equation} \label{eq:locality_PT}
\exp\left(-2\pi i\eta(\overline{X}) \right) = \exp\left(-2\pi i \int_{Y}\Phi_{d+2} \right)=\exp\left(-2\pi i\int_{\overline{X}} I_{d+1} \right) ,
\end{equation}
where in the last step we use Stokes' theorem. In a similar way, one can replace the $\eta$-invariant in the formula~\eqref{eq:inflow_Npert} by the Chern--Simons integral. In other words Eq.~\eqref{eq:inflow_Npert} is reduced to~\eqref{eq:inflow_pert} in perturbation theory. Furthermore, the right-hand side of~\eqref{eq:locality_PT} evaluates to the trivial phase on all boundaries $\overline{X}=\partial Y$ if and only if
\begin{equation}
\Phi_{d+2} = 0 ,
\end{equation}
recovering the usual condition for local anomaly cancellation.

Secondly, when the local anomalies cancel, {\em i.e.} when $\Phi_{d+2}=0$, 
the APS formula immediately implies that $\exp\left(-2\pi i\eta_{\overline{X}} \right)=1$ on all closed manifolds $\overline{X}$ that are boundaries of manifolds in one dimension higher, with structures extended. 
Thus $\exp\left(-2\pi i\eta_{\overline{X}} \right)$ becomes a bordism invariant~\cite{Witten:1985xe}, and so provides a homomorphism
\begin{equation}
\exp\left(-2\pi i\eta(\overline{X}) \right) \in \mathrm{Hom}(\Omega^{\mathrm{Spin}}_{d+1}(BG), U(1)),
\end{equation}
in the case that $\sigma$ is a {\em bona fide} spin structure.
Locality of the $d$-dimensional theory then requires this  be the trivial homomorphism. To find out whether this is the case for a given Dirac operator, bordism invariance means that our task is reduced to that of computing $\exp\left(-2\pi i\eta \right)$ on generators of $\Omega^{\mathrm{Spin}}_{d+1}(BG)$, which is typically some finite abelian group. Nonetheless, that task might still be a formidable one.

Of course, if the relevant spin bordism group $\Omega^{\mathrm{Spin}}_{d+1}(BG)$ vanishes, then the preceding arguments immediately imply that $\exp\left(-2\pi i\eta(\overline{X}) \right)=1$ on all closed $(d+1)$-manifolds (when $\Phi_{d+2}=0$). Hence, the pair of conditions
\begin{equation} \label{eq:2conditions}
\Phi_{d+2}=0 \quad \text{and} \quad \Omega^{\mathrm{Spin}}_{d+1}(BG)=0
\end{equation}
are sufficient for locality of the $d$-dimensional theory. Equivalently, these conditions are sufficient for the cancellation of local and global anomalies respectively. These criteria have recently been applied to investigate global anomalies in theories relevant to particle physics, for example in~\cite{Garcia-Etxebarria:2018ajm,Davighi:2019rcd,Wan:2019fxh}. The spin bordism group $\Omega^{\mathrm{Spin}}_{d+1}(BG)$ detects the most general possible global anomaly. 

In the next Section, we explain directly how global anomalies, as traditionally analyzed using mapping tori like the one discussed in the previous paragraph, are necessarily trivialized by enforcing the locality condition~\eqref{eq:locality}. Ultimately, locality is enough to guarantee cancellation of all known local and global anomalies. Indeed, enforcing the condition~\eqref{eq:locality} on the $\eta$-invariant precludes a big generalization of the possible ambiguities in the partition function that are detected by traditional mapping tori. This generalization of global anomalies (beyond the subset that are detected by mapping tori) has sometimes been referred to as `Dai--Freed anomaly cancellation'~\cite{Garcia-Etxebarria:2018ajm} due to the central role played by $\eta$-invariants and the gluing formula.

\section{Traditional global anomalies newly interpreted}\label{sec:trad}

We have seen that vanishing of the anomaly polynomial $\Phi_{d+2}$ only guarantees that $Z[A,M_d]$ is invariant under gauge transformations that are connected to the identity. We also saw that, from the anomaly inflow perspective, the condition $\Phi_{d+2}=0$ follows necessarily from locality of the $d$-dimensional theory.

The anomaly polynomial does not necessarily tell us about invariance of the partition function under gauge transformations that cannot be reached by infinitesimal ones. Suppose that spacetime has spherical topology (as is naturally motivated by taking a theory on flat space $\R^d$ and requiring the fields ``die off'' at infinite radius, allowing the point at infinity to be compactified). In this case, non-vanishing of the homotopy group $\pi_d(G)$ means there exists a gauge transformation $g:S^d \to G$, $[g] \neq 0 \in \pi_d(G)$, that cannot be connected to the identity by successive infinitesimal gauge transformations. This suggests there {\em could} be an anomaly, at least when the theory is defined on a sphere, arising from the gauge transformation $A\mapsto A^g = gAg^{-1} + i g \dd g^{-1}$ that cannot be explained by the local anomaly formula \eqref{eq:localanom}. 

To work out whether there is really an anomaly, one then has to analyze the spectral flow of the Dirac operator coupled to the background gauge field as one interpolates between $A$ and $A^g$ via a gauge field configuration in $d+1$ dimensions, for example by considering $A_t = (1-t)A + tA^g$, $t\in I:=[0,1]$.
This spectral flow can be used to deduce whether the partition function, which recall is a regularized product of the eigenvalues of the Dirac operator, is invariant under $A\to A^g$. If not there is a global anomaly. This was the original line of argument used to show that a 4d $SU(2)$ gauge theory with a single Weyl doublet is anomalous~\cite{Witten:1982fp}, because an odd number of eigenvalues of $i\slashed{D}^+$ change sign under the spectral flow for $[g]\in\pi_4(SU(2))$. Indeed, this spectral flow constraint was computed using 5d mod 2 index theorem of Atiyah and Singer~\cite{10.2307/1970757} on a mapping torus obtained by gluing together the ends of the interval $I$. Most saliently, this mod 2 index is a bordism invariant, meaning that the result of this computation only depends on the bordism class of the 5-manifold (equipped with $SU(2)$ connection and spin structure).

\begin{figure}[h]
  \centering
\begin{subfigure}[c]{0.4\textwidth}
    \centering
\includegraphics[scale=0.2]{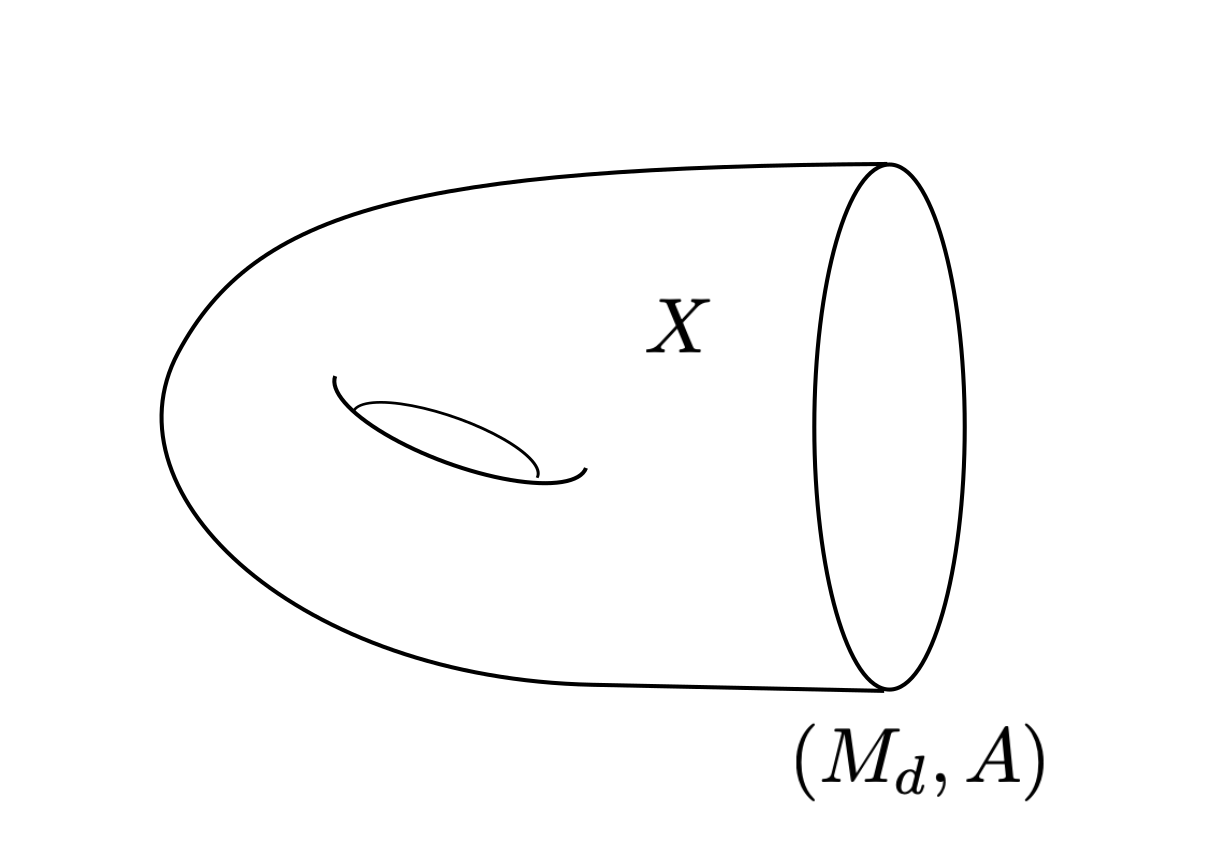}
\caption{}
\label{fig:MAext}
\end{subfigure}
\begin{subfigure}[c]{0.4\textwidth}
  \centering
  \includegraphics[scale=0.2]{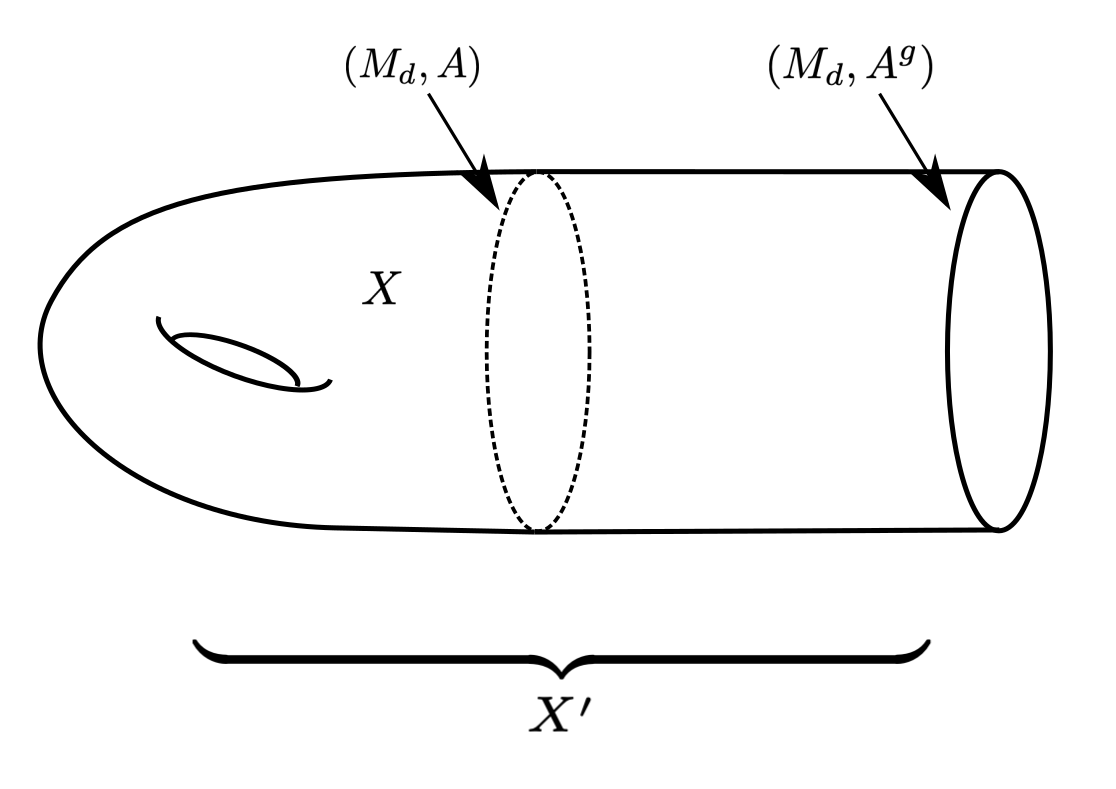}
  \caption{}
  \label{fig:MAgext}
  \end{subfigure}
  \caption{Extensions of (a) $(M_d,A)$ and (b) $(M_d, A^g)$ to $(d+1)$-manifolds.}
  \label{fig:TwoExt}
\end{figure}

This already hints at the underlying role played by bordism even in the original derivation of the $SU(2)$ global anomaly. Unsurprisingly, the traditional analysis for global anomalies outlined above can be recast, in the general case, using the non-perturbative anomaly inflow formula of \S\ref{sec:cobordism}, as follows. Suppose that we can extend the manifold $M_d$ with the background gauge field $A$, together with other structures needed to defined the theory such as the metric and the spin structure $\sigma$, to a $(d+1)$-dimensional manifold $X$ (see Fig.~\ref{fig:MAext}).\footnote{If this is not possible, then we can relate the theory on $M_d$ to a theory on a null-bordant manifold $\tilde{M}_d$, which belongs to the trivial class in $\Omega^{\Spin}_d(BG)$, by a gravitational theta-angle, and continue the analysis on $\tilde{M}_d$~\cite{Freed:2004yc,Witten:2019bou}.} Then the putative fermionic partition function $Z[A,M_d]$ is, by anomaly inflow,
\begin{equation}
  \label{eq:putativeZA}
  Z[A,M_d] = \left|Z[A,M_d]\right| \exp\left(-2\pi i \eta(X,{\bf R})\right),
\end{equation}
where we now emphasize that the evaluation of the $\eta$-invariant depends on the representation of the fermions to which $A$ is coupled. Anomaly inflow also allows us to write down a formula for the partition function evaluated (for the same $M_d$) on the background $A^g$, which recall is related to $A$ by the homotopically non-trivial gauge transformation. To wit, one needs to form an appropriate bounding $(d+1)$-manifold, call it $X^\prime$, which can be done by gluing $X$ to a cylinder $M_d\times [0,1]$ with an interpolating background gauge field $A_t = (1-t)A + tA^g$, as illustrated in Fig.~\ref{fig:MAgext}. The partition function for this background can then be written as
\begin{equation}
  \label{eq:putativeZA}
  Z[A^g,M_d] = \left|Z[A,M_d]\right| \exp\left(-2\pi i \eta (X^\prime,{\bf R})\right),
\end{equation}
where we have used the fact that the modulus of the partition function is necessarily anomaly-free to write $\left|Z[A^g,M_d]\right| = \left|Z[A,M_d]\right|$. The fermionic partition function therefore varies at most by a phase, as we know on general grounds.

\subsection{Mapping tori}

The gluing formula \eqref{eq:glue} for the exponentiated $\eta$-invariant then allows one to compare $Z[A^g]$ and $Z[A]$ by expressing their quotient in terms of the $\eta$-invariant evaluated on the interpolating cylinder, since
\begin{equation}
  \label{eq:etaMappingcylinder}
  \frac{Z[A^g,M_d]}{Z[A,M_d]} = \frac{\exp\left(-2\pi i \eta (X^\prime,{\bf R})\right)}{\exp\left(-2\pi i  \eta (X,{\bf R})\right)} = \exp\left(-2\pi i \eta (M_d\times [0,1],{\bf R})\right).
\end{equation}
One can then identify the two boundary components $M_d\times \{0\}$ and $M_d\times \{1\}$ of the interpolating cylinder as usual, because $A$ and $A^g$ are gauge equivalent. The result is a closed $(d+1)$-manifold $\overline{X}_g$ equipped with a certain $G$-bundle called a mapping torus, which as a manifold is $\overline{X}_g = M_d\times S^1$. Using the gluing formula once more, this implies the variation of the partition function is computed by the $\eta$-invariant on the mapping torus,
\begin{equation}
  \frac{Z[A^g,M_d]}{Z[A,M_d]} = \exp(-2\pi i \eta (\overline{X}_g,{\bf R})).\label{eq:mappingTanom}
\end{equation}
The partition function is then well-defined on gauge fields modulo this gauge transformation by $g$ only if
\begin{equation}\label{eq:anomcancel}
  \exp(-2\pi i \eta (\overline{X}_g, {\bf R})) = 1.
\end{equation}
If this phase is not equal to one, then the homotopically non-trivial gauge transformation by $g$ is anomalous. But note that in the derivation above we have not used the fact that the gauge transformation $g$ cannot be smoothly connected to the identity, so the expression \eqref{eq:mappingTanom} is in fact valid for any gauge transformation. For any $g \in G$, regardless of its homotopy class in $\pi_d(G)$, the condition \eqref{eq:anomcancel} is clearly implied by the much more general locality condition \eqref{eq:locality}. And since this locality is guaranteed by the pair of conditions in Eq. \eqref{eq:2conditions}, these bordism-based conditions are sufficient for cancellation of all anomalies, local or global, regardless of $\pi_d(G)$. In other words, global anomalies are detected by `Omega' rather than `pi'. Nonetheless, $\pi_d(G)$ does encode partial information about the global anomaly, in the case that spacetime is a sphere, as we discuss next.

\subsection{Mapping spheres and the role of pi}

When the original spacetime manifold $M_d$ is topologically a $d$-sphere $S^d$, one can arguably simplify the problem of computing the global anomaly by working with a $(d+1)$-dimensional sphere as our closed $(d+1)$-manifold $\overline{X}$, instead of the mapping torus described above. We refer to this as a \emph{mapping sphere}, constructed as follows. 

\begin{figure}[h]
  \centering
\begin{subfigure}[c]{0.8\textwidth}
    \centering
\includegraphics[scale=0.35]{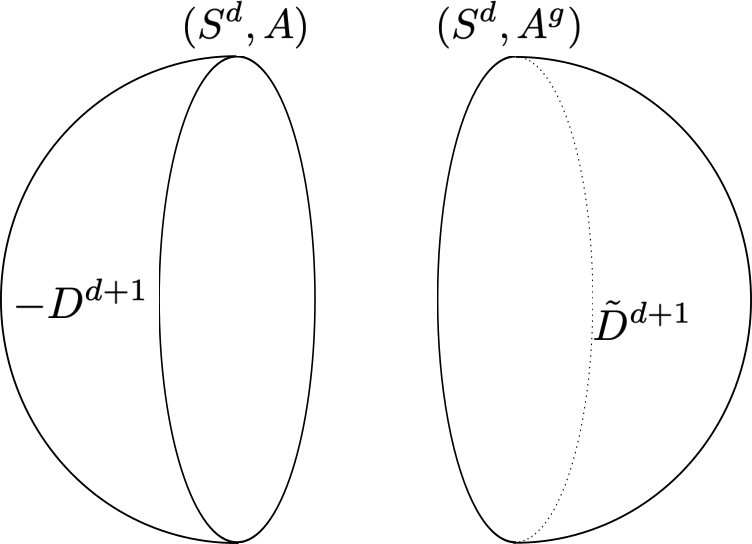}
\end{subfigure}

\begin{subfigure}[c]{0.8\textwidth}
  \centering
  \includegraphics[scale=0.35]{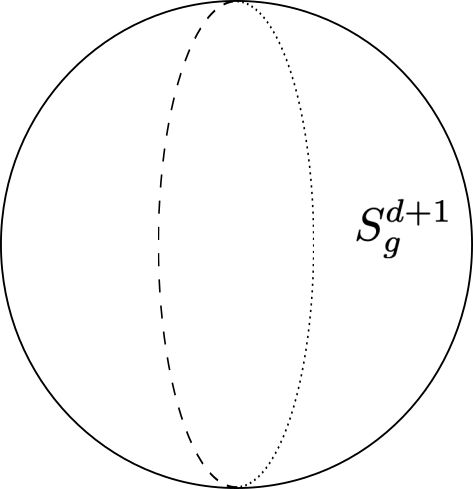}
\end{subfigure}
\caption{Gluing of two hemispheres $\tilde{D}^{d+1}$ and $-D^{d+1}$ with a shared boundary $S^d$ to form a mapping sphere $S^{d+1}_g$.} \label{fig:mappingsphere}
\end{figure}

Firstly, since one should detect an anomaly under $A \to A^g$ for any choice of background $A$ on $S^d$, let us choose $A$ to define a trivial bundle on $S^d$ to simplify the analysis. The trivial bundle on $S^d$ can then always be extended to a trivial bundle on a hemisphere $D^{d+1}$ that $S^d$ bounds. Similarly, since $A^g$ is gauge equivalent to $A$, it must also define a trivial bundle on $S^d$ which can also be extended to a trivial bundle on another hemisphere, call it $\tilde{D}^{d+1}$. Repeating the procedure in the previous subsection, one can use the gluing properties of the $\eta$-invariant to write
  \begin{equation}
    \label{eq:mappingS}
    \frac{Z[A^g,S^d]}{Z[A,S^d]} = \exp\left(-2\pi i \eta(S^{d+1}_g, {\bf R})\right),
  \end{equation}
  where the mapping sphere $S^{d+1}_g$ is constructed by gluing $\tilde{D}^{d+1}$ and $-D^{d+1}$ along the $S^d$ boundary with the gauge transformation $g$, as shown in Fig.~\ref{fig:mappingsphere}. Such a construction is also used in Ref.~\cite{Lee:2020}.

With this construction in hand, the precise role of the homotopy group $\pi_d(G)$ in global anomaly cancellation becomes more apparent. The bundle on the mapping sphere $S^{d+1}_g$ corresponds to the homotopy class of a map $f:S^{d+1}_g \to BG$. This map can be lifted to a ball whose boundary is the mapping sphere precisely when $\pi_{d+1}(BG)$ vanishes. When $\pi_{d+1}(BG)$ is non-vanishing, there exist bundles on the mapping sphere which cannot be extended to a bulk $(d+2)$-ball. There is a natural isomorphism
\begin{equation}
\pi_{d+1}(BG) \cong \pi_d(G)
\end{equation}
when $G$ is a compact Lie group, which means that the bundle on the mapping sphere is non-trivial precisely when the two hemispheres are connected using a gauge transformation in the non-trivial class of $\pi_d(G)$. Hence, non-vanishing of the group $\pi_d(G)$ means there are mapping spheres which cannot be extended to bulk balls, for which the anomalous phase of the partition function could thence be evaluated purely in terms of the local anomaly polynomial $\Phi_{d+2}$. This is the precise obstruction encoded by $\pi_d(G)$, as described in the Introduction. However, crucially, this is not enough to establish that there is a global anomaly, because the APS index theorem can be used when the mapping sphere is the boundary of {\em any} bulk manifold with structures extended, not necessarily a $(d+2)$-ball. We develop this point further in our discussion of 6d $SU(2)$ anomalies in \S \ref{sec:no_global}.

\subsection*{Back to the $SU(2)$ anomaly}

Let us return to the example of an $SU(2)$ gauge theory in four dimensions. Here $\Phi_{6}=0$ so there is no perturbative anomaly. Hence, the exponentiated $\eta$-invariant is a cobordism invariant, and any residual global anomaly will be measured by evaluating the $\eta$-invariant on generators of the bordism group, here $\Omega^\Spin_5(BSU(2)) \cong \Z/2$. In this case, the mapping torus constructed above (by gluing with a gauge transformation with $[g]\in \pi_4(SU(2))$) is a suitable generator of $\Omega^\Spin_5(BSU(2)) \cong \Z/2$, and we have recalled above how the 5d mod 2 index theorem can be used to evaluate the variation of the partition function via spectral flow. 

Bordism invariance means that one can compute the global anomaly induced by this large gauge transformation by evaluating $\exp 2\pi i \eta$ on {\em any} manifold in the same bordism class as that original mapping torus, and there are arguably simpler choices available. One alternative choice is the mapping sphere, $S^5$ with non-trivial bundle, as we described for this 4d $SU(2)$ example in the Introduction. For a second alternative, which is arguably the simplest to work with, consider the closed manifold $S^4 \times S^1$ equipped with an $SU(2)$ connection with unit instanton number through the $S^4$ factor (that is constant around the $S^1$), and equipped with a spin structure with periodic boundary conditions around the $S^1$ factor. This choice of spin structure corresponds to a mapping torus whose ends are glued together using $(-1)^F$ which is equivalent to a constant $SU(2)$ gauge transformation. This alternative mapping torus, which must yield the same global anomaly by bordism invariance, was discussed in~\cite{Wang:2018qoy}. 

The exponentiated $\eta$-invariant can thence be computed by a variety of methods, for example using the 5d mod 2 index theorem again. Alternatively, one can embed the $SU(2)$ connection inside a large group whose 5$\textsuperscript{th}$ spin bordism group vanishes, such as $U(2)$ or $SU(3)$, and then extend the bundle to a bounding 6-manifold and use the APS index theorem to relate $\eta$ to the integral of a degree-6 anomaly polynomial~\cite{Davighi:2020bvi,Davighi:2020uab}.
For the isospin $j$ representation of $SU(2)$, the calculation yields
\begin{equation}
  \exp(-2\pi i \eta (S^4\times S^1, {\bf R}_j)) =(-1)^{T(j)}, \quad  T(j) = \frac{2}{3}j(j+1)(2j+1)
\end{equation}
Therefore, there can be at most a mod $2$ global anomaly for the chiral fermion in the isospin $j$ representation depending on the parity of $T(j)$. Since $T(j)$ is odd if and only if $j\in2\Z+1/2$, only chiral fermions in these representations contribute to the mod $2$ global anomaly.

\subsection{$\pi_d \neq 0$ is not necessary}\label{sec:notnec}

To continue our comparison of the role of bordism {\em vs.} homotopy groups in detecting global anomalies, we now emphasize that non-vanishing of $\pi_d(G)$ is neither necessary nor sufficient for a $d$-dimensional gauge theory, with gauge group $G$, to exhibit a global anomaly (for any choice of fermion content).

Firstly, $\pi_d(G) \neq 0$ is not necessary for a theory to exhibit a global anomaly. 
Over the years, more subtle global anomalies that cannot be accounted for by the homotopy groups have been discovered. It is clear to see why the homotopy group method fails in these cases. The homotopy group only tells us whether about the possibility for smooth gauge transformations on a Euclidean space that approach the identity transformation as $|x|\to \infty$. The condition $\pi_d(G) \neq 0$ is then necessary for the possibility of a global anomaly in the `vanilla' situation where $G$ is a continuous gauge group, spacetime is taken to be a sphere, and there are no extra structures that play a significant role. This can be seen by considering extendability of the mapping sphere construction to an interior ball, as we explained in the Introduction. But, when one or more of these conditions do not hold, the condition $\pi_d(G) \neq 0$ is no longer necessary for there to be a global anomaly, as the following simple examples illustrate.

\paragraph{Example 1: discrete gauge groups.}

Consider a theory with a discrete gauge group $G = \Z/k$, in any spacetime dimension. No local anomaly can arise because there are no infinitesimal gauge transformations. But there are in general global anomalies, as can be seen for example from \emph{e.g.} $\Omega^{\Spin}_3(B\Z/2) \cong \Z/8$, corresponding to a mod 8 valued anomaly in a 2d theory with a unitary $\Z/2$ symmetry~\cite{Davighi:2020uab,Ryu:2012he,Kaidi:2019pzj,Kaidi:2019tyf}. However, this anomaly cannot technically be seen from the homotopy group $\pi_2(\Z/2)$, which vanishes because any base-point preserving map must map the whole $S^2$ to a fixed base point in $\Z/2$ by continuity. Discrete gauge anomalies in four dimensions were first studied in Refs.~\cite{Ibanez:1991hv,Ibanez:1991pr}, and were revisited from the bordism perspective in~\cite{Hsieh:2018ifc}.

\paragraph{Example 2: parity anomaly.}

Non-trivial global anomalies can also arise when 
there is a non-trivial interplay between the gauge group and the spacetime symmetry. In such cases one cannot hope to use the homotopy group of the gauge group to detect a global anomaly since by definition it cannot see anomalies on a topologically non-trivial spacetime, and it knows nothing about any extra structures that might be present in the theory's definition.

An example of this failure of the homotopy-based condition comes from the `parity' anomaly on unorientable manifolds~\cite{Witten:2016cio}, as follows. Consider a $U(1)$ gauge theory in $3$ dimensions with a charge $1$ massless Dirac fermion on an unorientable manifold equipped with a $\Pin^+$ structure. The theory suffers from a global anomaly as can be seen from the non-trivial exponentiated $\eta$-invariant evaluated on closed unorientable $3$-manifolds. The anomaly persists even after including two Majorana fermions that transform under time-reversal with the opposite sign to cancel the mod $16$ gravitational anomaly \cite{Hsieh:2015xaa,Witten:2015aba}. By an explicit evaluation of the $\eta$-invariant on $\R P^3$, Witten found in Ref.~\cite{Witten:2016cio} that the anomaly is of order $4$, that is, we need $4$ copies of charge-1 Dirac fermions (as well as an appropriate number for Majorana fermions to cancel the gravitational anomaly) for the theory to be completely anomaly-free. One can also see this mod $4$ anomaly detected in the bordism group $\Omega^{\Pin+}_4(BU(1)) \cong \Z/4\times \Z/16$, whose computation is presented in Appendix~\ref{app:Wittenparity}. The anomaly can be interpreted as a mixed anomaly between time-reversal symmetry and the $U(1)$. However, contrary to the usual `parity' anomaly in $3$d \cite{Redlich:1983kn,Redlich:1983dv} where the existence of the mixed anomaly simply implies time-reversal symmetry is broken (as we must uphold the gauge symmetry for the theory to be well-defined), here we have `gauged' the time-reversal symmetry in the sense that we used it to define the theory on unorientable manifolds and so the presence of the mixed anomaly must mean that the theory cannot be defined consistently.

\paragraph{Example 3: beyond spheres.}

Even without adding extra structures like a time-reversal symmetry, and even if we stick to `vanilla' continuous gauge groups, the homotopy condition is unlikely to be necessary for global anomalies if we are interested in theories defined on arbitrary spacetime manifolds, not just spheres. And one should always be interested in such a setup, since locality ultimately requires that a well-defined quantum field theory should be defined in such generality.

To illustrate this possibility we are content to contrast Omega {\em vs.} pi in an example. Consider a theory in 8 spacetime dimensions, with gauge group $G=E_7$ or $E_8$. In either case, the relevant homotopy groups vanish, $\pi_8(E_7)=\pi_8(E_8)=0$ (see {\em e.g.}~\cite{Garcia-Etxebarria:2017crf}). However, the relevant spin-bordism groups do not, with~\cite{Garcia-Etxebarria:2018ajm} 
\begin{equation}
\Omega^{\text{Spin}}_9(BE_7)\cong\Omega^{\text{Spin}}_9(BE_8)\cong\Z/2.
\end{equation}
This suggests these theories may indeed suffer from a global anomaly when evaluated on some 8-manifold which is not a sphere. The $E_8$ theory in particular would be interesting to investigate, since it might have implications for heterotic string theory compactified down to eight dimensions.

\subsection{$\pi_d \neq 0$ is not sufficient}\label{sec:notsuf}

On the other hand, suppose that $\Omega^\Spin_{d+1}(BG) = 0$, so the mapping torus $X$ necessarily bounds a $(d+1)$-dimensional manifold $Y$ with all structures appropriately extended. Even if $\pi_d(G)\neq 0$, we have by virtue of the APS index theorem that
\begin{equation}
  \label{eq:APS}
  \eta(X,{\bf R}) = \int_Y \Phi_{d+2}\quad \mod \Z,
\end{equation}
where $\Phi_{d+2}$ is the anomaly polynomial \eqref{eq:anom-poly}. This means that, for any fermion content such that there is no local anomaly, there can be no possible non-perturbative anomaly. Indeed, we have already seen that, even in the case where $G$ is connected and $M_d$ is a sphere, a non-trivial homotopy group does not guarantee there can be a global anomaly, because the mapping sphere could still be filled in by some bulk $(d+2)$-manifold that is not a ball on which the APS index theorem can be used.

It is also worth emphasising once more that it is possible that $Z[A^g] \neq Z[A]$ under a gauge transformation $g$ in the non-trivial homotopy class of $\pi_d(G)$, even when the bordism group vanishes. However, this would just correspond to a local anomaly, since the phase shift is completely given in terms of the anomaly polynomial. In such a case, considering homotopically non-trivial gauge transformations cannot give rise to constraints in addition to the local anomaly cancellation conditions. An important class of examples with $\pi_d(G) \neq 0$ but no possible global anomaly is provided by theories in 6d. We analyze these theories in detail in \S \ref{sec:6dexamples}.

\paragraph{Example: $U(2)$ gauge theory in 4d.}

A simple example to illustrate insufficiency of $\pi_d \neq 0$ for global anomalies is provided by $U(2)$ gauge theory in $4$ dimensions. Just as is the case for the gauge group $SU(2)$, the fourth homotopy group of $U(2)$ is $\pi_4(U(2)) \cong \Z/2$, and one can take the homotopically non-trivial gauge transformation $g(x)$ to be the same as in the $SU(2)$ gauge group by embedding $SU(2)$ in $U(2)$. However, it can be shown that $\Omega^\Spin_5(BU(2))$ vanishes~\cite{Davighi:2019rcd,Wan:2019fxh}, signalling yet another conflict between Omega and pi. 

Accordingly, the `alternative' mapping torus that we described above for $SU(2)$, which is $S^4\times S^1$ equipped with a 1-instanton $SU(2)$ background on the $S^4$ factor and a $(-1)^F$ twist (implemented via a periodic spin structure) on the $S^1$ factor, can be extended to a 6-manifold bounded by it by embedding $SU(2)$ as a subgroup of $U(2)$. The crucial point is that the twist by $(-1)^F$ can be equivalently realised as the holonomy of a $U(1)\subset U(2)$-component of the connection around the $S^1$ direction. Parametrizing the $S^1$ direction by a local coordinate $\phi\in [0,2\pi)$, a suitable such choice of $U(2)$ connection is
\begin{equation}
  A_\phi = \frac{1}{2}\dd\phi {\bf 1}_2 + A_{\text{inst}},
\end{equation}
where $A_{\text{inst}}$ is the original $1$-instanton configuration on $S^4$. Now that we can take the spin structure to be anti-periodic around the $S^1$ it can be extended to a hemisphere $H^2$ that $S^1$ bounds, parametrized by coordinates $(\theta,\phi)$ with $\theta \in [0,\pi/2]$. The gauge field configuration can also be extended, because the $\dd\phi$ contribution coincides with the gauge field configuration of a charge-2 $U(1)$ monopole at the centre of a unit $S^2$ when restricted to its equator. 

Thus, we see that this 5-dimensional $U(2)$ mapping torus (and indeed any other) can always be extended to a 6-manifold that it bounds, and so the exponentiated $\eta$-invariant can always be evaluated via the APS index theorem. Thence, if the fermionic matter content is such that the theory is free of local anomalies, there can be no further anomaly. Indeed, the easiest way to see this is to evaluate the mixed gauge-gravitational anomaly coefficient modulo $2$. The vanishing of this anomaly coefficient implies that the difference between the number of left-handed and right-handed fermions with isospins $j\in 2\Z_{\geq 0}+1/2$ under $SU(2)\subset U(2)$ (which are the only representations contributing to the $\Z/2$-valued global anomaly in $SU(2)$) must be even, as discussed extensively in Ref. \cite{Davighi:2020bvi}.

\section{Anomaly cancellation in 6 dimensions}\label{sec:6dexamples}

In 6 dimensions, the homotopy group $\pi_6(G)$ vanishes for most simple Lie groups $G$, apart from when $G=SU(2)$, $SU(3)$, or $G_2$. As a result it has been widely reported that these theories all feature possible global anomalies (see {\em e.g.} Refs.~\cite{Kiritsis:1986mf,Bershadsky:1997sb,Dobrescu:2001ae,Suzuki:2005vu}). Cancelling these global anomalies has been linked, for example, to the existence of three generations of Standard Model fermions in theories with two extra dimensions~\cite{Dobrescu:2001ae}.

However, the spin bordism groups of the corresponding classifying spaces $BG$ in degree $7$ all vanish for these gauge groups, as compared side by side in Table~\ref{tab:BG2bord}. The bordism group results for $SU(2)$ and $SU(3)$ have been recently computed in Refs.~\cite{Garcia-Etxebarria:2018ajm,Monnier:2018nfs}. To complete the picture, we show that the $7$\textsuperscript{th} spin bordism group of $BG_2$ also vanishes in Appendix~\ref{app:G2bord}, using the Adams spectral sequence.
\begin{table}[h]
  \centering
  \begin{tabular}{|c|c|c|}
    \hline
    $G$ & $\Omega^{\Spin}_7(BG)$ & $\pi_6(G)$ \\
    \hline
    $SU(2)$ & $0$ & $\Z/12$ \\
    $SU(3)$ & $0$ & $\Z/6$ \\
    $G_2$ & $0$ & $\Z/3$\\
    \hline
  \end{tabular}
  \caption{$\Omega$ \emph{versus} $\pi$ for $G=SU(2)$, $SU(3)$, and $G_2$.}\label{tab:omegavspi}
\end{table}

\subsection{No global anomalies} \label{sec:no_global}

The vanishing of these bordism groups means, as always, that there can be no possible global anomaly in any of these 6d gauge theories, because the exponentiated $\eta$-invariant on any closed 7-manifold (with any consistent $G$-bundle and spin structure) can always be evaluated from the local anomaly polynomial $\Phi_8$ using the APS index theorem. Knowing this to be the case, it is nonetheless instructive to carefully examine {\em why} the specific mapping tori whose ends are glued using homotopically non-trivial gauge transformations do not lead to global anomalies, as was previously thought to be the case~\cite{Kiritsis:1986mf,Bershadsky:1997sb,Dobrescu:2001ae,Suzuki:2005vu}. To that end we next seek to explain, albeit somewhat schematically, how the 7d mapping torus can be explicitly realised as the boundary of a suitable 8-manifold in the case of $G=SU(2)$.

In order to highlight the difference between using homotopy groups and bordism groups, it is especially instructive to fix the spacetime topology to be $S^6$ and work with mapping spheres rather than the more traditional mapping tori. As explained in Section~\ref{sec:trad}, a mapping sphere is constructed from two hemispherical halves, each of which is equipped with a trivial $SU(2)$ bundle, but which are glued together at their equators using a gauge transformation $g(x)\in SU(2)$ that cannot be smoothly deformed to the identity. For example, let $g$ have homotopy class $[g(x)]=1\mod 12 \in \pi_6(SU(2))=\Z/12$. The gauge bundle on $S^7$ constructed in this way naturally lives in the class $1\mod 12 \in \pi_7(BSU(2)) \cong \pi_6(SU(2))$. Recalling that $G$-bundles on $X$ are classified (up to isomorphism) by homotopy classes of maps from $X$ to $BG$, it follows that $\pi_7(BSU(2))$ classifies $SU(2)$ bundles on $S^7$. Thus, the bundle on the mapping sphere is topologically non-trivial precisely when the two hemispheres are glued together using a homotopically non-trivial gauge transformation.

Since the bundle on $S^7$ is non-trivial, we cannot thence extend it to an 8-ball $B^8$ filling the mapping sphere. This is the `obstruction' that is properly detected by the non-vanishing homotopy group $\pi_6(SU(2))$. But this obstruction does not mean that there is a global anomaly. Rather, there can be a global anomaly only if there is an obstruction that prevents us from simultaneously extending both the gauge bundle and the spin structure from the mapping sphere to {\em any} bulk 8-manifold. The homotopy calculation simply tells us that we have to search beyond just 8-balls, and rather consider bulks that are non-contractible.

To find such a bulk, or at least to demonstrate that one exists, one might first think to exploit the homology of $BSU(2)$. Because the homology group $H_7(BSU(2);\Z)$ vanishes, there always exists an 8-chain in $BSU(2)$ whose boundary is the 7-cycle defined by a map $f:S^7\to BSU(2)$. However, at this point our homological line of attack comes unstuck, because such a chain cannot necessarily be realised as an embedded submanifold of $BSU(2)$, as we would need in order to use the APS index theorem. (Indeed, once the degree $q$ exceeds 6, as is the case here, Thom showed~\cite{thom1954quelques} that there exist homology classes which are not representable by the image under any smooth map -- not necessarily even an embedding -- of the fundamental class of a closed $q$-manifold.\footnote{We thank Ben Gripaios and Yuji Tachikawa for discussions related to this point.})
In fact, this `shortcoming' of homology was a primary motivation for the development of bordism theory in the first place, and so at this point in our discussion it is appropriate to pass to bordism. Fortunately, as one can view bordism groups as a generalised homology theory, there are various algebraic methods for computing them, such as the Atiyah--Hirzebruch spectral sequence (AHSS). Once the spin bordism groups $\Omega^\Spin_\bullet(\text{pt})$ of a point are known, one can use the simple fibration $\text{pt}\to BSU(2)\to BSU(2)$ to construct the spectral sequence.\footnote{See {\em e.g.} Refs.\cite{Monnier:2018nfs,Garcia-Etxebarria:2018ajm,Davighi:2019rcd} for more details on the AHSS computations in related contexts.} 
  
To compute the spin bordism group $\Omega^\Spin_{d+1}(BG)$ with the AHSS, one builds successive stacks (commonly called pages) of complexes $\{E^r_{p,q}\}_{r\geq 2}$ of abelian groups, where these pages are equipped with group homomorphisms called `differentials', $\dd_r: E^r_{p,q} \rightarrow E^r_{p-r,q+r-1}$. Each element $E^r_{p,q}$ on page $r$ gives an approximation to $\Omega^\Spin_{p+q}(BG)$, starting from the second page
\begin{equation}
  E^2_{p,q} = H_p(BG; \Omega^\Spin_q(\text{pt})). \label{eq:AHSSE2}
\end{equation}
 The approximation is refined from one page to the next by taking a homology of the complex on the previous page. More precisely,
\begin{equation}
  E^{r+1}_{p,q} = \frac{\text{ker}\,\left[\dd_r: E^r_{p,q} \rightarrow E^r_{p-r,q+r-1}\right]}{\text{im}\,\left[\dd_r: E^r_{p+r,q-r+1} \rightarrow E^r_{p,q}\right]}.
\end{equation}
For $G=SU(2)$, the second page factorizes into a tensor product because there is no torsion in $BSU(2)$, {\em viz.} 
$$E^2_{p,7-p} = H_p(BSU(2);\Z)\otimes \Omega^\Spin_{7-p}(\text{pt}).$$
This tells us about whether we can simultaneously extend `gauge bundles on $p$-cycles' to $(p+1)$-chains in $BSU(2)$ that they bound (ignoring spin structures), and spin structures on $(7-p)$-manifolds to bulk $(8-p)$-manifolds (with trivial gauge bundle).

In our particular case of $d=6$ and $G=SU(2)$, the extension of any gauge bundle on $S^7$ to an 8-chain in $BSU(2)$ (ignoring the existence of a spin structure) corresponds to the element $E^{2}_{7,0} = 0$, while the extension of any spin structure on $S^7$ (this time ignoring the gauge bundle) is captured by $E^2_{0,7} = 0$. However, this is just a crude approximation to our extendability problem in terms of certain cycles and chains. One then needs to examine the behaviour of smaller complementary cycles, encoded in the elements $E^2_{p,7-p}$, to get a more complete picture (albeit still in terms of chains). Ultimately, it is the vanishing of all the other $E^2_{p,7-p}$ elements that tells us that such a chain can be realised as a submanifold and there is no obstruction to a simultaneous extension of any gauge bundle and spin structure on $S^7$ to a spin 8-manifold. This is what is needed to defer to the APS index theorem, and thus show that any anomalous phase is captured by the local anomaly polynomial. 
\begin{figure}[h]
\centering
\begin{subfigure}[b]{0.45\textwidth}
  \centering
  \includegraphics[width=1\textwidth]{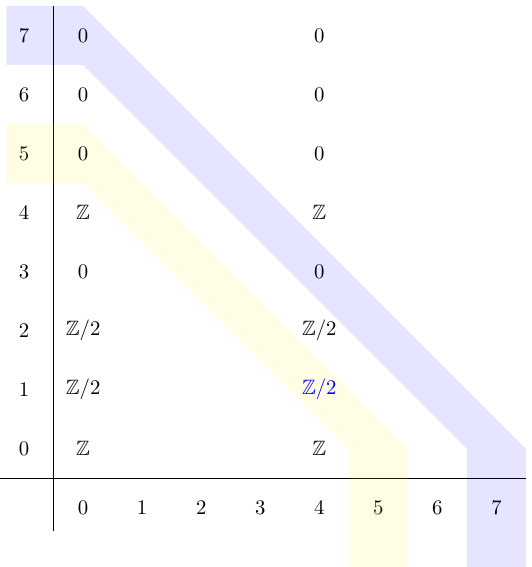}
   \caption{}
   \label{fig:AHSSSU2} 
 \end{subfigure}
\begin{subfigure}[b]{0.45\textwidth}
  \centering
  \includegraphics[width=1\textwidth]{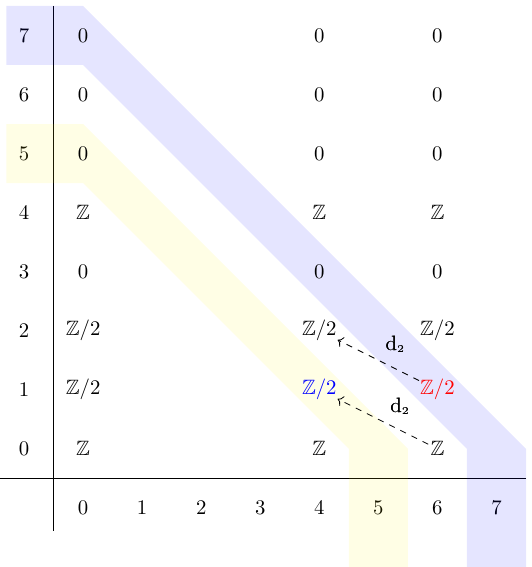}
   \caption{}
   \label{fig:AHSSSU3}
\end{subfigure}
\caption{The second page of the AHSS for {\bf (a)} $\Omega^\Spin_\bullet(BSU(2))$ and {\bf (b)} $\Omega^\Spin_\bullet(BSU(3))$} \label{fig:AHSS}
\end{figure}

It is worth noting that the situation is a little different when the gauge group is $SU(3)$, both in 4 and 6 dimensions, which we now comment on only briefly. The `first approximation' to the bordism group by $E^2_{p,q}$, which splits the extendability of the spin structure and the extendability of the gauge bundle into complementary $p$- and $q$-cycles, is no longer trivially vanishing as it was for $G=SU(2)$ (in degree 7). Even though $H_7(BSU(3);\Z) = 0$, we thus cannot deduce from the second page of the AHSS that an 8-chain to which the $SU(3)$-bundle on $S^7$ extends is realisable as a submanifold of $BSU(3)$.
But this first approximation now gets refined due to non-vanishing differentials on the second page, which take into account how the spin structure and the gauge bundle intertwine (as shown in Fig.~\ref{fig:AHSSSU3}). The coloured elements do indeed vanish on the next page, meaning that there is no obstruction to simultaneously extending both an $SU(3)$ gauge bundle and a spin structure to some 8-manifold that $S^7$ (or indeed any other closed 7-manifold, such as a traditional mapping torus) bounds.

In principle, one can write the anomalies $\exp(-2\pi i \eta(S^7_g))$ in terms of the anomaly polynomial evaluated on the extended spin 8-manifold $M_8$. In practice, however, one might not be able to construct such an extension explicitly. One can view the traditional anomaly interplay method of Elitzur and Nair \cite{Elitzur:1984kr} (see \emph{e.g.} \S3 of Ref.~\cite{Lee:2020} for a treatment more closely aligned to our current perspective) as one way to circumvent this computational problem. 

\subsection{Simple local anomaly cancellation conditions}\label{sec:6dlocal}

It has been emphasized that in order to analyze global anomalies one must first cancel the local ones. We now discuss how the latter can be achieved for completeness. We restrict attention to the simplest scenario whereby there are only contributions to the local anomaly from spin-$1/2$ fermions. Thus, consider a matter content of Weyl fermions in the representation ${\bf R} = {\bf R}_L \oplus {\bf R}_R$ of the gauge group $G$. Those fermions in the representation ${\bf R}_L$ are left-handed and those in the ${\bf R}_R$ representation are right-handed. We can further decompose these representations in terms of irreducible representations as
\begin{equation}
  {\bf R}_L = \bigoplus_i {\bf r}_{iL},\qquad   {\bf R}_R = \bigoplus_i {\bf r}_{iR}.
\end{equation}
The simplest scenario when these anomalies cancel is when the anomaly polynomial $\Phi_8(A;{\bf R}) = \Phi_8(A;{\bf R}_L)-\Phi_8(A;{\bf R}_R)$ vanishes. The anomaly polynomial for the left-handed representation ${\bf R}_L$ can be written explicitly as
\begin{equation}\label{eq:8danompoly}
  \Phi_8(A;{\bf R}_L) = \frac{1}{4! (2\pi)^4} \tr_{{\bf R}_L} F^4 + \frac{1}{48 (2\pi)^2} p_1 \tr_{{\bf R}_L} F^2 + \frac{1}{5760}(7p_1^2-4p_2) \text{dim} ({\bf R}_L),
\end{equation}
where $p_1$ and $p_2$ are the first and the second Pontryagin classes of the tangent bundle, respectively. The anomaly polynomial for the right-handed representation ${\bf R}_R$ is given by the same formula with ${\bf R}_L$ replaced by ${\bf R}_R$.

For a finite dimensional irreducible representation ${\bf r}$ of the gauge group, we can write
\begin{equation}
  \tr_{{\bf r}} F^2 = C_2({\bf r};G) \tr_{{\bf f}} F^2,
\end{equation}
where $C_2({\bf r};G)$ is the second order Casimir invariant of the gauge group $G$ in the representation ${\bf r}$, and  ${\bf f}$ is the non-trivial irreducible representation with lowest dimension. Moreover, for any finite dimensional irreducible representation ${\bf r}$ of the gauge groups $G=SU(2),\; SU(3)$, and $G_2$, the fourth order Casimir invariants vanish, and we can write
\begin{equation}
  \tr_{{\bf r}}F^4 = C_4({\bf r}; G)(\tr_{{\bf f}} F^2)^2,
\end{equation}
where $C_4({\bf r};G)$ is given in terms of the second order Casimir invariants, as well as the dimensions of the group $G$ and the representation ${\bf r}$, by~\cite{Okubo:1978qe,Okubo:1981td}
\begin{equation}
  C_4({\bf r};G) = \frac{\text{dim}\,G}{2\,\text{dim}\,{\bf r} (2+\text{dim}\,G)}\left(6-\frac{C_2({\bf ad};G)}{\text{dim}\,G} \frac{\text{dim}\,{\bf r}}{C_2({\bf r};G)}\right) C_2({\bf r};G)^2.
\end{equation}
Since the three terms in the expression \eqref{eq:8danompoly} are independent, we can rewrite the anomaly cancellation conditions in terms of group-theoretic quantities as
\begin{align}
  \sum_i C_4({\bf r}_{iL};G) - \sum_i C_4({\bf r}_{iR};G) &= 0,\label{eq:C4cond}\\
  \sum_i C_2({\bf r}_{iL};G) - \sum_i C_2({\bf r}_{iR};G) &= 0,\label{eq:C2cond}\\
  \sum_i\text{dim}\,{\bf r}_{iL} - \sum_i \text{dim}\,{\bf r}_{iR} &=0.\label{eq:dimcond}
\end{align}
When these conditions are satisfied, our analysis in \S\ref{sec:notsuf} and the results from Table \ref{tab:omegavspi} tell us that there can be no anomaly arising from the gauge transformations that cannot be smoothly connected to the identity.

From a quick comparison of these conditions and the conditions for local anomaly cancellation in the literature, we can clearly see that the anomaly polynomial in most papers on anomaly cancellation in 6d apparently does not vanish, in which case we cannot proceed to discuss global anomalies. However, there is no real contradiction here, for the following two reasons. Firstly, in the analysis above we have not taken into account the contribution from other fields such as the self-dual 2-form gauge field or the spin-$3/2$ Rarita-Schwinger fields that are ubiquitous in 6d supergravity, which is often the setting of the discussion on this topic. Secondly, unlike in four dimensions, local anomaly cancellation in 6d does not require the anomaly polynomial to vanish identically. One can cancel the anomalies through a counterterm involving one or more self-dual 2-form gauge fields. We turn to this alternative route of local anomaly cancellation in the next Subsection.

\subsection{The Green--Schwarz mechanism}\label{eq:6dgreenschwarz}

It is well known that the problem of anomaly cancellation has a cohomological flavour to it. It is not necessary for the anomaly polynomial to vanish identically for the theory to be anomaly-free; if one can add counterterms to absorb the gauge non-invariance of the original anomalous partition function, then the resulting partition function is made gauge invariant and hence anomaly-free. If the total anomaly polynomial factorizes, one can add to the original action a counterterm constructed from a self-dual 2-form gauge field and the characteristic classes of the gauge and tangent bundles to cancel the local anomalies. This method of cancelling perturbative anomalies, known as the \emph{Green--Schwarz mechanism}\cite{Green:1984sg}, is especially important in supersymmetric theories where there are contributions to the anomaly from higher-spin fields, because supersymmetry severely restricts the matter content such that one cannot get an anomaly-free theory by a vanishing anomaly polynomial without making the theory completely trivial. 

It is in this supersymmetric context that the 6d `global' anomalies due to gauge transformations in a non-trivial homotopy class of $\pi_6(G)$ are often discussed (see Refs. \cite{Bershadsky:1997sb,Suzuki:2005vu} for some examples). The anomaly cancellation conditions given in these references are derived ultimately from local anomaly cancellation conditions, possibly with the aid of the Green--Schwarz mechanism just described, by embedding $G$ as a subgroup of some larger group whose $6^\mathrm{th}$ homotopy group vanishes. This indirect method for deriving an anomaly associated with a homotopically non-trivial gauge transformation was pioneered by Elitzur and Nair in Ref. \cite{Elitzur:1984kr}. But, crucially, there is no guarantee that such an anomaly is a global one, in the sense that it persists even when the local anomaly vanishes.
Indeed, when the bordism group $\Omega^\Spin_7(BG)$ vanishes, we know that there can be no global anomaly. Therefore, the 6d `global' anomaly cancellation conditions, derived from homotopically non-trivial gauge transformations, 
in fact just provide necessary conditions for the local anomalies to be cancelled by the Green--Schwarz mechanism. There is no additional global anomaly.

This claim was explicitly proven recently by Monnier and Moore for $\mathcal{N}=(1,0)$ 6d supergravity theories~\cite{Monnier:2018nfs}. In such theories, one can cancel the local anomalies by a generalization of the Green--Schwarz mechanism if the total anomaly polynomial factorizes as  
\begin{equation}
  \Phi^{(tot)}_8 = \frac{1}{2} Y \wedge Y,\label{eq:anomfactors}
\end{equation}
where $Y$ is an appropriately quantized closed $4$-form constructed from the gauge field strength $F$ and the Riemann curvature 2-form $R$. Then the generalised Green--Schwarz counterterm modifies the partition function by
\begin{equation}
  Z_{\text{ct}}[Y,B,M_6] = \exp\left(2\pi i \int_{M_6} \frac{1}{2} B\wedge Y\right), \label{eq:GSct}
\end{equation}
where $B$ is a self-dual 2-form gauge field. The variation of $Z_{\text{ct}}[Y,B,M_6]$  under infinitesimal gauge transformations and diffeomorphisms precisely cancels the perturbative anomalies of the original action when we impose the modified Bianchi identity \cite{Green:1984bx,Akbulut:2016aP,Sadov:1996zm,Riccioni:2001bg} 
\begin{equation}
  \dd H = Y,
\end{equation}
where $H$ is the 3-form field strength (or curvature form) associated to the 2-form gauge field ({\em i.e.} 2-form connection) $B$. Note that we have expressed the Green--Schwarz counterterm \eqref{eq:GSct} in terms of a local 2-form gauge field, so the formula is valid only locally. To construct a version that works globally, one can write the Green--Schwarz counterterm \eqref{eq:GSct} as a \emph{shifted Wu--Chern--Simons} theory~\cite{Monnier:2016jlo} in 7d, which is a generalisation of a 3d `spin Chern--Simons theory' \cite{Dijkgraaf:1989pz,Jenquin:2005,Jenquin:2006jh,Belov:2005ze}. To cancel the anomalies, the shifted Wu--Chern--Simons theory must be isomorphic to the conjugated anomaly theory that describes the anomalies of our anomalous 6d supergravity theory. In Ref. \cite{Monnier:2018nfs}, Monnier and Moore showed that the two theories are isomorphic up to a bordism invariant. In particular, if $\Omega^\Spin_7(BG)$ vanishes then all anomalies can be cancelled by the Green--Schwarz mechanism (in the form of the shifted Wu--Chern--Simons theory), provided the anomaly polynomial factorizes as in~\eqref{eq:anomfactors}. 
Therefore, if one can cancel the perturbative anomalies via the Green--Schwarz mechanism, one cannot have further anomalies due to the homotopically non-trivial gauge transformations probed by $\pi_6(G)$.

\subsection*{Acknowledgments}
We thank Pietro Benetti Genolini, Philip Boyle Smith, Ben Gripaios, Avner Karasik, John March-Russell, David Tong, and Carl Turner for helpful discussions. We especially thank Yuji Tachikawa and Yasunori Lee for their many helpful suggestions which improved this manuscript. NL is supported by David Tong’s Simons Investigator Award. We are supported by the STFC consolidated grant ST/P000681/1.

\appendix
\section{Some Bordism Group Calculations}\label{app:bordismcal}

In this Appendix we present two calculations of relevant bordism groups mentioned in the text that we believe are new. We employ the Adams spectral sequence in both cases. For a guide to the Adams spectral sequence aimed at physicists, we recommend especially Ref.\cite{Campbell:2017khc,beaudry2018guide}. The present authors also offered a condensed explanation of the method in \cite{Davighi:2020uab}. Unless stated explicitly otherwise, all cohomology groups in this appendix have coefficients in $\Z/2$.

\subsection{$\Omega^{\Pin+}_4(BU(1))$}\label{app:Wittenparity}

Let us start by computing the bordism group $\Omega^{\Pin+}_4(BU(1))$,
which captures 't Hooft anomalies between a $U(1)$ gauge symmetry and
time-reversal symmetry $T$ with $T^2 = (-1)^F$. This will correspond to an anomaly
when we put a fermionic gauge theory with gauge group $U(1)$ on an
unorientable manifold that admits a $\Pin^+$ structure.

Here the relevant stable symmetry type is $H = \Pin^+\times U(1)$. The
corresponding Madsen-Tillmann spectrum is given by
\begin{equation}
  \label{eq:MTPinU1}
  MT(\Pin^+\times U(1)) = M\Spin \wedge \Sigma MTO_1 \wedge BU(1)_+ , 
\end{equation}
where the subscript $+$ denotes a disjoint base point. Since we are
interested in the $4$\textsuperscript{th} degree, the
Anderson--Brown--Peterson Theorem applies and we have the Adams
spectral sequence
\begin{equation}
  E^{s,t}_2 = \text{Ext}^{s,t}_{\mathcal{A}_2(1)}\left(\Sigma H^\bullet(MTO_1)\otimes H^\bullet(BU(1)),\Z/2\right) \Longrightarrow \Omega_{t-s}^{\Pin+}(BU(1))^{\wedge}_2
\end{equation}
for the $2$-completion of the bordism group, where $\mathcal{A}_2(1)$ is the subalgebra of the Steenrod algebra $\mathcal{A}_2$ generated by $\sq^1$ and $\sq^2$.

\begin{figure}[h!]
   \centering
    \begin{subfigure}[t]{0.5\textwidth}
    \centering
 \begin{tikzpicture}[scale=.5] 
   \fill (0, 0) circle (3pt) node[anchor=north]  {$\scriptstyle \bar{U}$};
   \fill (0, 1) circle (3pt) node[anchor=west] {$\scriptstyle w_1\bar{U}$};
   \fill (0, 2) circle (3pt) node[anchor=east] {$\scriptstyle w_1^2\bar{U}$};
   \fill (0, 3) circle (3pt) node[anchor=west] {$\scriptstyle w_1^3\bar{U}$};
   \fill (0, 4) circle (3pt) node[anchor=west] {$\scriptstyle w_1^4\bar{U}$};
   \fill (0,5) circle (3pt) node[anchor=east] {$\scriptstyle w_1^5\bar{U}$};
   \fill (0,6) circle (3pt) node[anchor=east] {$\scriptstyle w_1^6\bar{U}$};
   \fill (0,7) circle (3pt) node[anchor=east] {$\scriptstyle w_1^7\bar{U}$};
   \fill (0,8) circle (3pt) ;
   \fill (0,9) circle (3pt);
   \fill (0,10) circle (3pt);
   \sqone (0,0, black);
   \sqtwoL (0,0, black);
   \sqone (0,2, black);
   \sqtwoL (0,3, black);
   \sqone (0,4, black);
   \sqtwoR (0,4, black);
   \sqone (0,6, black);
   \sqtwoR (0,7, black);
   \sqtwoL (0,8, black);
   \sqone (0,8, black);
   \sqone (0,10, dashed);

    \fill (6, 0+2) circle (3pt) node[anchor=north]  {$\scriptstyle c_1\bar{U}$};
   \fill (6, 1+2) circle (3pt) node[anchor=west] {$\scriptstyle c_1w_1\bar{U}$};
   \fill (6, 2+2) circle (3pt) node[anchor=east] {$\scriptstyle (c_1^2+c_1w_1^2)\bar{U}$};
   \fill (6, 3+2) circle (3pt);
   \fill (6, 4+2) circle (3pt);
   \fill (6,5+2) circle (3pt);
   \fill (6,6+2) circle (3pt);
   \fill (6,7+2) circle (3pt);
   \fill (6,8+2) circle (3pt) ;
   \fill (6,9+2) circle (3pt);
   \fill (6,10+2) circle (3pt);
   \sqone (6,0+2, black);
   \sqtwoL (6,0+2, black);
   \sqone (6,2+2, black);
   \sqtwoL (6,3+2, black);
   \sqone (6,4+2, black);
   \sqtwoR (6,4+2, black);
   \sqone (6,6+2, black);
   \sqtwoR (6,7+2, black);
   \sqtwoL (6,8+2, black);
   \sqone (6,8+2, black);
   \sqone (6,10+2, dashed);

   \fill (8, 4) circle (3pt) node[anchor=west] {$\scriptstyle c_1^2\bar{U}$};
   \fill (8,5) circle (3pt);
   \fill (8,6) circle (3pt);
   \fill (8,7) circle (3pt);
   \fill (8,8) circle (3pt) ;
   \fill (8,9) circle (3pt);
   \fill (8,10) circle (3pt);
   
   \sqone (8,4, black);
   \sqtwoR (8,4, black);
   \sqone (8,6, black);
   \sqtwoR (8,7, black);
   \sqtwoL (8,8, black);
   \sqone (8,8, black);
   \sqone (8,10, dashed);
   \sqtwoCR (6,3, black);
\end{tikzpicture}
\caption{}
\label{fig:HMTOU1module}
\end{subfigure}
\begin{subfigure}[t]{0.4\textwidth}
    \centering
\includegraphics[scale=0.7]{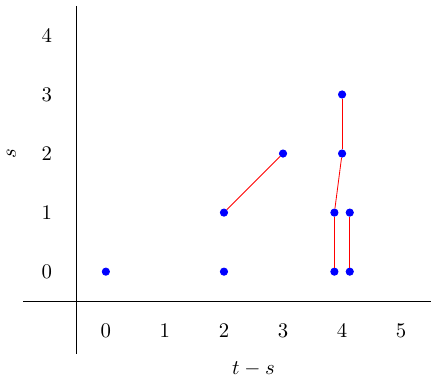}
\caption{}
\label{fig:Adamsparity}
\end{subfigure}
\caption{{\bf (a)} The $\mathcal{A}_2(1)$-module structure for $\Sigma H^\bullet(MTO_1)\otimes H^\bullet(BU(1))$ and {\bf (b)} its corresponding Adams chart.}
\label{fig:HMTOU1}
\end{figure}
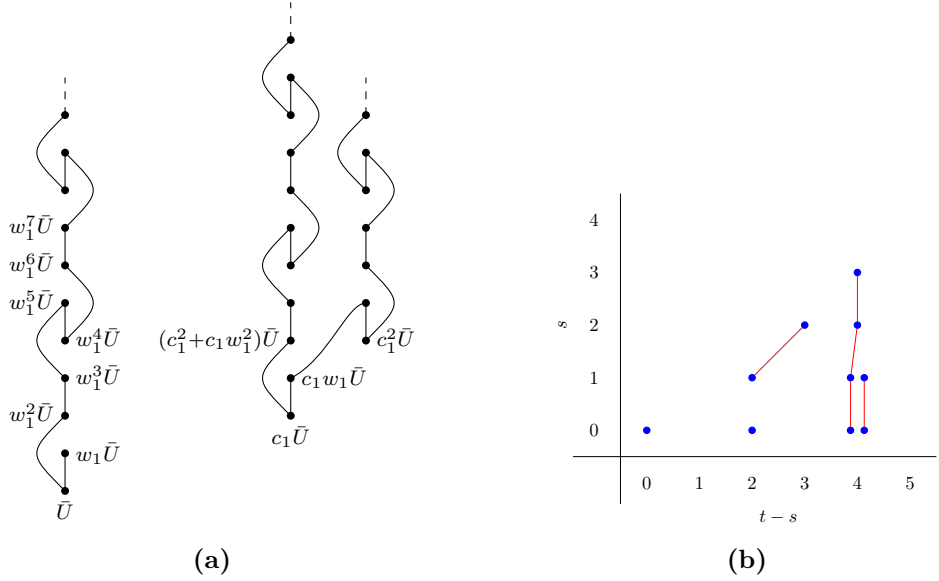

It is well known that
\begin{align}
  H^\bullet(MTO_1) &\cong \Z/2[w_1]\{\bar{U}\},\\
  H^\bullet(BU(1)) &\cong \Z/2[c_1],
\end{align}
where $w_1$ is the first Stiefel--Whitney class of $O(1)$,
$\bar{U}= U(-\gamma_1) \in H^{-1}(MTO_1)$ is the Thom class evaluated
on the virtual bundle $-\gamma_1$ of the universal bundle
$\gamma_1$. $c_1$ is of course the first Chern class of $U(1)$. Thus we have
\begin{equation}
  H^\bullet(MTO_1)\otimes H^\bullet (BU(1)) \cong \Z/2[w_1,c_1]\{\bar{U}\}.
\end{equation}
The $\mathcal{A}_2(1)$-module structure of
$H^\bullet(MTO_1)\otimes H^\bullet (BU(1))$ can be determined by the
action of the Steenrod squares $\sq^1$ and $\sq^2$ on the
generators. The Steenrod square actions on the Thom class $\bar{U}$ is
given by $\sq^i \bar{U} = \bar{w}_i \bar{U}$ where $\bar{w}_i$ can be
defined in terms of the total Stiefel--Whitney class of the virtual
bundle $-\gamma_1$ as 
\begin{equation}
  \bar{w} = w(-\gamma_1) = 1+\sum_{i\geq 1} \bar{w}_i.
\end{equation}
But we know that
$w(-\gamma_1) = \frac{1}{w} = 1 + w_1^2 + w_1^3 + \ldots$, whence
$\bar{w}_i = w_1^i$. Therefore, the action of the Steenrod squares of
$\bar{U}$ can be written in the form
\begin{equation}
  \sq^i \bar{U} = w_1^i \bar{U}.
\end{equation}
The Steenrod square actions on $w_1$ and $c_1$ are given by
\begin{align}
  \label{eq:Wuformula}
  \sq^1 w_1 = w_1^2, &\quad \sq^2 w_1 =0,\\
  \sq^1 c_1 = 0, &\quad \sq^2 c_1 = c_1^2.
\end{align}
One can subsequently work out the $\mathcal{A}_2(1)$-module structure
of $H^\bullet(MTO_1)\otimes H^\bullet (BU(1))$, as shown in
Fig.~\ref{fig:HMTOU1module}, with the associated Adams chart given in
Fig.~\ref{fig:Adamsparity}.

We can then read off from the Adams chart that
\begin{equation}
  \label{eq:Omegapin+4}
  \Omega^{\Pin+}_4(BU(1)) \cong \Z/4 \times \Z/16.
\end{equation}

\subsection{$\Omega^{\Spin}_7(BG_2)$}\label{app:G2bord}
\begin{figure}[h!]
   \centering
    \begin{subfigure}[t]{0.2\textwidth}
    \centering
 \begin{tikzpicture}[scale=.5] 
   \fill (0, 0) circle (3pt) node[anchor=north]  {$\scriptstyle 1$};
   \fill (0, 4) circle (3pt) node[anchor=east] {$x_4$};
   \fill (0, 6) circle (3pt) node[anchor=east] {$x_6$};
   \fill (0, 7) circle (3pt) node[anchor=east] {$x_7$};
   \sqtwoR (0,4,black);
   \sqone (0,6,black);
\end{tikzpicture}

\caption{}
\label{fig:HBG2module}
\end{subfigure}
\begin{subfigure}[t]{0.5\textwidth}
    \centering
\includegraphics[scale=0.7]{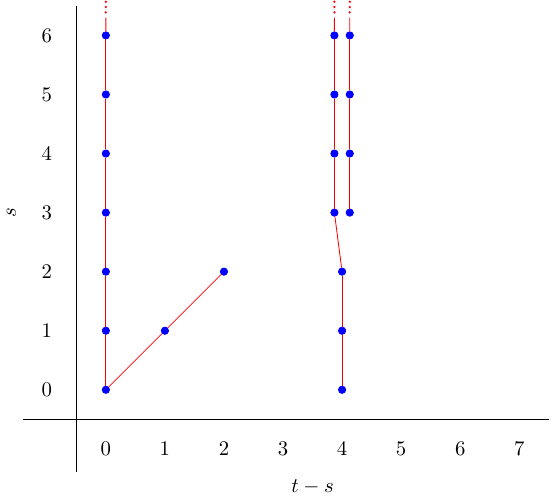}
\caption{}
\label{fig:AdamsBG2}
\end{subfigure}
\caption{{\bf (a)} The $\mathcal{A}_2(1)$-module structure for  $H^\bullet(BG_2)$ and {\bf (b)} its corresponding Adams chart.}
\label{fig:HBG2}
\end{figure}
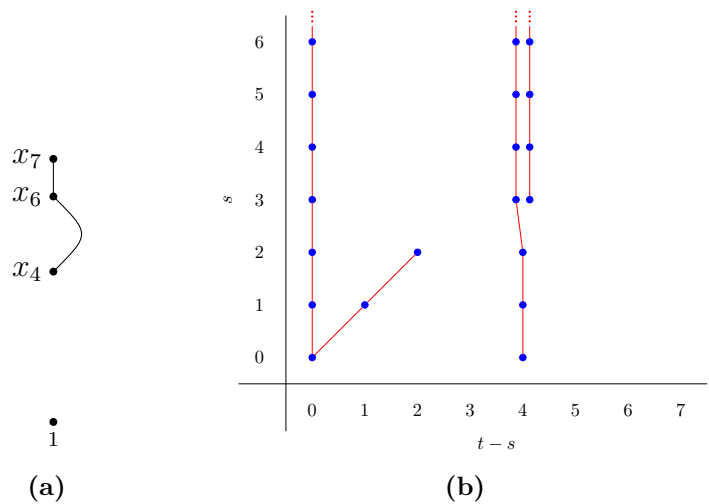

We can use the same procedure to calculate $\Omega_7^{\Spin}(BG_2)$
for the exceptional Lie group $G_2$. Since we are interested in the
$7$\textsuperscript{th} degree, the Anderson--Brown--Peterson Theorem
still applies and we still have the Adams spectral sequence
\begin{equation}
  E^{s,t}_2 = \text{Ext}^{s,t}_{\mathcal{A}_2(1)}\left(H^\bullet(BG_2),\Z/2\right) \Longrightarrow \Omega_{t-s}^{\Spin}(BG_2)^{\wedge}_2
\end{equation}
for the $2$-completion of the bordism group.  We know that
$H^\bullet(BG_2) = \Z/2[x_4,x_6,x_7]$ with the non-trivial actions of
$\mathcal{A}_2(1)$ given by $ \sq^2x_4 = x_6, \sq^1x_6 = x_7$. The
$\mathcal{A}_2(1)$-module structure of $H^\bullet(BG_2)$ and the
corresponding Adams chart is shown in Fig. \ref{fig:HBG2}, and we
readily find that $\Omega^{\Spin}_7(BG_2)^{\wedge}_2 = 0$. To complete
the calculation, we still need to show that there is no odd torsion
involved.  Theorem 2.19 of Ref.~\cite{Akbulut:2016aP} computed the
integral cohomology of $BG_2$ to be free of odd torsion:
\begin{equation}
  H^\bullet(BG_2;\Z) \cong \Z[y_4,y_{12}] \oplus \Z/2[y_6,y_{10}],
\end{equation}
with $\text{deg}~y_k = k$. Hence the integral homology of $BG_2$ is also
devoid of odd torsion by the universal coefficient theorem. The
Atiyah--Hirzebruch spectral sequence from the fibration
$\text{pt}\rightarrow BG_2 \rightarrow BG_2$ that computes the spin
bordism groups of $BG_2$ is given by
\begin{equation}
  E^2_{p,q} = H_p(BG_2;\Omega^{\Spin}_q(\text{pt})) \Longrightarrow \Omega^{\Spin}_{p+q}(BG_2).
\end{equation}
As there is no odd torsion in the spin bordism group of a point in
degrees less than $8$, the relevant elements in the second page cannot
contain any odd torsion, and thus cannot give rise to any odd torsion
in the $7$\textsuperscript{th} spin bordism groups for $BG_2$. The
spin bordism groups for $BG_2$ up to degree $7$ are given in
Table~\ref{tab:BG2bord}.
\begin{table}[h!]
  \centering
  \begin{tabular}{c|cccccccc}
    $d$ & $0$ & $1$ & $2$ & $3$ & $4$ & $5$ & $6$ & $7$ \\
    \hline
    $\Omega^{\Spin}_d(BG_2)$ & $\Z$ & $\Z/2$ & $\Z/2$ & $0$ & $\Z\times \Z$ & $0$ & $0$ & $0$
  \end{tabular}
  \caption[$BG_2$ spin bordism]{Spin bordism groups for $BG_2$ up to degree $7$}
  \label{tab:BG2bord}
\end{table}

\newpage
\bibliography{references}
\bibliographystyle{JHEP}

\end{document}